# PULSATIONS, BOUNDARY LAYERS, AND PERIOD BOUNCE

# IN THE CATACLYSMIC VARIABLE RE J1255+266

JOSEPH PATTERSON,[1] JOHN R. THORSTENSEN,[2] & JONATHAN KEMP[3,1]

[1] Department of Astronomy, Columbia University, 550 West 120th Street, New York, NY 10027; jop@astro.columbia.edu

[2] Department of Physics and Astronomy, Dartmouth College, 6127 Wilder Laboratory, Hanover, NH 03755; j.thorstensen@dartmouth.edu

[3] Joint Astronomy Centre, University Park, 660 North A'ohōkū Place, Hilo, HI 96720; j.kemp@jach.hawaii.edu

## ABSTRACT

We report time-series photometry of the quiescent optical counterpart of the EUV transient RE J1255+266. The star appears as a DA white dwarf with bright emission lines and a complex spectrum of periodic signals in its light curve. A signal at 0.0829 d is likely to be the orbital period of the underlying cataclysmic binary (probably a dwarf nova). Characteristic periods of 1344, 1236, and 668 s are seen, and a host of weaker signals. We interpret these noncommensurate signals as (nonradial) pulsation periods of the white dwarf. The donor star is unseen at all wavelengths, and the accretion rate is very low. We estimate a distance of 180±50 pc, and $M_v$=14.6±1.3 for the accretion light. The binary probably represents a very late stage of evolution, with the donor star whittled down to $M_2$<0.05 $M_\odot$. Such binaries stubbornly resist discovery due to their faintness and reluctance to erupt, but are probably a very common type of cataclysmic variable. If the signal at 0.0829 d is indeed the orbital period, then the binary is an excellent candidate as a "period bouncer". Plausible colleagues in this club include four dwarf novae and the (so far) noneruptive stars GD 552 and 1RXS J105010.3–140431 (hereafter RX 1050).

The 1994 EUV eruption implies a soft X-ray/EUV luminosity of $10^{34-35}$ erg/s, greater than that of any other dwarf nova. We attribute that to a favorable blend of properties: a high-mass white dwarf; a very transparent line of sight (the "Local Chimney"); and a low binary inclination. The first maximizes the expected temperature and luminosity of boundary-layer emission; the other two increase the likelihood that soft X-rays can survive their perilous passage through an accretion-disk wind and the interstellar medium.

*Subject headings*: accretion, accretion disks — binaries: close — novae, cataclysmic variables — stars: individual (RE J1255+266)

## 1. INTRODUCTION

RE J1255+266 (hereafter abbreviated as RE 1255, and sometimes informally called "Watson X–1") was discovered as a very bright EUV transient by the ROSAT Wide-Field Camera, and then optically identified by Watson et al. (1996, hereafter W96) as a DA white dwarf with emission lines. The transient was quite remarkable; it was the only such event ever seen in the 6-month sky survey, and at maximum light became the brightest object in the EUV sky. Dahlem et al. (1995, see also the discovery announcement by Dahlem & Kreysing 1994) estimated a peak flux of $7 \times 10^{-9}$ ergs/cm$^2$/s in the 112–200 Å band. W96 interpreted this as an outburst of a previously unknown dwarf nova, although clear evidence of binary motion was lacking.

A long EUVE observation in 1995 found no source at this position, implying a brightening by a factor >50000 (Drake et al. 1998). Wheatley et al. (2000) found a marginal hard X-ray source, tending to confirm the W96 interpreation as a cataclysmic variable (CV). If this is correct, then the limits on binary motion suggest a star of extreme mass ratio, possibly a very old CV. Such stars have great significance in the theory of binary evolution, but are very difficult to discover. This led us to undertake observations of the quiescent counterpart, which we report here.

## 2. OBSERVATIONS

The photometry was obtained with the CCD photometers at the 1.3 m and 2.4 m telecopes of MDM Observatory. The observing log is shown in Table 1. Most of the time-series observations used a wide 4000–8000 Å bandpass filter, which improved signal-to-noise but prevented calibration to any standard magnitude system. On several occasions we used standard *UBVRI* filters, yielding calibrated magnitudes.

### *2.1 ASTROMETRY*

Some of the images were obtained for astrometric purposes, as part of a parallax program. Measured J2000 coordinates were RA = 12h 55m 10.57s, Dec = +26d 42m 26.8s (±0.3"), Comparing a 2003 January 30 image with six images from the USNO Plate Measuring Machine scans (back to 1955), we found a proper motion

$$\mu_x = -27(6) \text{ mas/yr}$$
$$\mu_y = -12(6) \text{ mas/yr},$$

where the systematic errors are probably comparable to the random error. This agrees with the result from the USNO–B1.0 catalog (–32 and –8 mas/yr respectively), but is probably more accurate since it includes the recent very precise CCD image.

The proper motion enables a (very) rough estimate of distance. Adopting our favorite model (Thorstensen 2003) for the velocity ellipsoid of CVs, this implies a median distance of 160 pc, with a 1σ low at 80 and a 1σ high at 280 pc (16- and 84-percentile points of an asymmetric distribution). This compares acceptably to ~200 pc which we will estimate below from the white-dwarf evidence, especially the pulsations.

## 2.2 TIME–SERIES PHOTOMETRY AND NIGHTLY POWER SPECTRA

The time-series observations consist of 26 nights of differential photometry with respect to a nearby comparison star (USNO 1125–06614414, with $B$=19.7, $R$=18.2). Because we varied the wide-band filters, we generally could not establish the nightly average magnitude any more accurately than ~0.4 mag; within that large uncertainty, the star did not vary. More interestingly, the light curve on individual nights was extremely quiet, never varying more than 0.1 mag. Two sample light curves are shown in the upper frames of Figure 1. That near-constancy is unusual; most CVs flicker erratically by much greater amounts. The wideband measurements suggest an equivalent $R$ magnitude of 19.5, but this is uncertain by ~0.4 due to the impossibility of obtaining calibration with such wide filters. Snapshot calibrated magnitudes indicated $R$=19.1.

We calculated the power spectrum of each night's light curve. Obvious periodic signals appeared in "Run 2", shown in Figure 1 with the dates labelled (JD 270 and 271). The frequencies were 69, 138, and 207 (each ±2) c/d. Three months later, clear signals were seen on each of the seven nights comprising Run 3. These occurred near 64 and 129 (±0.7) c/d, and are shown in the lower frames of Figure 1. The amplitudes varied from night to night, as suggested by Figure 1. A weak signal near 200 c/d may also have been detected on two nights (JD 378 and 382).

For most observations of short duration or low quality, no periodic signals were found. But the semi-amplitude upper limit on such nights was typically ~0.03 mag, so this was not quite good enough to provide a useful comparison to the nights of high quality. Detected periodic signals on individual nights are listed in Table 2A. The main features of that table are as follows.

1. For the two runs (2 and 3) with strong detections, the signals definitely appeared at different frequency (70/139 and 64/129 c/d respectively).

2. For two nights of Run 1, the periodic signals seem to have occurred at the frequencies found in Run 3.

3. There was a sprinkling of weak detections at frequencies seen just once. These should be regarded merely as candidates, awaiting possible confirmation.

4. The Nyquist frequency for many of these observations was ~1000 c/d, so the candidate signals at high frequency may well be aliased.

Run 3 provided by far the best coverage. We formed the average power spectrum of these seven nights (the "incoherent sum") and show the result in Figure 2, where the uncertainty in frequency is 0.7 c/d. The obvious signals near 64 and 129 c/d appear, and another feature at 11.4 c/d, which we will soon identify as the orbital frequency of the underlying binary. In order of increasing frequency, we will subsequently refer to these signals as $\omega_0$, $\omega_1$, and $\omega_2$.

### *2.3 POWER SPECTRA OF MERGED LIGHT CURVES*

A power spectrum of a long time series (over a few nights) is more sensitive to low-amplitude signals, and also gives much better frequency resolution. We calculated these power spectra for Runs 1, 2, 3, and 7, with significant detections identified in Table 2B. All except Run 3 show an annoying ±1 c/d alias problem; clarification of these signals (if real) awaits better measurement.

Run 2 was of special interest, because those signals were the strongest we ever detected (up to 0.04 mag semi-amplitude). The power spectrum of the two-night light curve is seen in Figure 3. We tested the hypothesis that the three frequencies are in an exact 1:2:3 ratio. The result was inconclusive, of course (for such short observations, nightly aliasing does not permit unambiguous choice of cycle count). But if the frequencies are integer multiples, there is only one viable fundamental: 69.90±0.10 c/d. We will call the three detections $\omega_3$, $\omega_4$, and $\omega_5$, in order of increasing frequency. But we emphasize that we have only measured these frequencies to ±2.4 c/d, and do not know whether $\omega_4$ and $\omega_5$ are independent of $\omega_3$ (or merely harmonics). Their entries in Table 2 reflect this ambiguity.

The coverage in Run 3 was of sufficient quality and density to overcome aliasing, and therefore warranted a much closer look. The relevant portions of the power spectrum of this 7-night time series are shown in Figure 4. At low frequencies, a weak signal at 12.07(2) c/d appeared. This is likely to signify the orbital frequency of the underlying binary. The only other strong and obvious signals occurred at 64.27(2) and 129.28(2) c/d. These are seen in the lower frames of Figure 4, along with the power spectrum window (for a signal at precisely 64.27 c/d, sampled like the actual data). Comparison of the actual power spectra with that of the window shows that most of the fine structure in Figure 3 arises from the windowing of the time series. The data are roughly consistent with the hypothesis of just these three signals.

But not *exactly* consistent. Close inspection of the power spectra suggests some excess power nestled within the picket-fence structure and not present in the window. To study this further, we "cleaned" the time series by subtracting the best-fit $\omega_0$, $\omega_1$, and $\omega_2$ sinusoids. The power spectrum of the residual time series (not shown) revealed broad bumps in the vicinity of $\omega_1$ and $\omega_2$. Closer analysis suggested that these probably arise from additional periodic signals hovering near the detection limit, although we could not exclude an alternative origin in amplitude/frequency variability of $\omega_1$ and $\omega_2$. Just for the record, the strongest of these lesser peaks occurred at 64.96/65.99 and 130.11/129.11 c/d. These candidate signals are too weak to survive the combined perils of aliasing and noise, so we cannot tell them apart. But the preferred choices are 64.96 and 130.11 (both ±0.02) c/d, slightly blue-shifted from $\omega_1$ and $\omega_2$.

### *2.4 ANOTHER LOOK AT LOW FREQUENCY*

We studied the low-frequency regime carefully, trying to discredit the hypothesis that 12.07 c/d is the orbital frequency (since this is potentially of great significance, as we will discuss below). We did not find any disqualifying evidence. That signal is of low amplitude, just 0.028 mag full amplitude. But upon removing a sinusoid of that amplitude, we found (in the power spectrum) no features of significance in its vicinity — consistent with the hypothesis of a steady signal. No other feature in a plausible frequency range appears with full amplitude greater than 0.018 mag. This does not prove the point, of course; maybe the true $\omega_{orb}$ is lost somewhere else in the low-frequency noise of Figure 4. This is an important possibility to check, preferably with an appropriate radial-velocity search. In the meantime, we adopt $\omega_{orb}$=12.07 c/d as a working hypothesis.

*2.5 STANDARD MAGNITUDES AND COLORS*

Table 1 also shows *UBVI* magnitudes obtained during 2001–4. No believable variability was seen during these snapshot observations, and they were obtained in close succession to make the color information more reliable. Lack of contemporaneous spectra prevented an accurate correction for emission-line contamination. The uncorrected colors, interpreted as those of a DA white-dwarf photosphere, imply that $T_{eff}$ is near 13000 K (Koester et al. 1979, Fontaine et al. 1982, Terashita & Matsushima 1966).

The squares in Figure 5 show the flux distribution. The *UBVI* magnitudes were obtained in 2003, and the *JHK* magnitudes were obtained in April 1995, 10 months after outburst. The continuous curve shows the flux distribution (on the scale at right) of G 226–29, a well-observed pulsating white dwarf with $T_{eff}$=12300 K and log $g$=8.3 (Bergeron et al. 1995). This curve is scaled to align the *V* flux exactly with that of RE 1255. Evidently it gives a fairly good fit to the *UBVI* fluxes of RE 1255. However, log $g$ and the unknown small corrections for emission lines contribute extra uncertainty. Allowing for those uncertainties in RE 1255, we estimate $T_{eff}$ in the range 11000–15000 K.

In April 1995 W96 measured *V*~18.0, which is 1.0 mag above what we now identify as "quiescence". Such a long-term decline years after outburst is characteristic of the WZ Sge class, to which this star clearly belongs. It is perhaps a major reason why the 1995 *JHK* fluxes appear so high in Figure 5. However, even after subtracting a suitably scaled-up white-dwarf flux, it seems likely that some extra infrared emission (*IJHK*) still exists.

**3. WHITE–DWARF PULSATIONS, DISTANCE, ACCRETION RATE, MASS RATIO**

The spectrum and flux distribution demonstrate that a DA white dwarf dominates the light at visual wavelengths. Because the white dwarf dominates, and because the high-frequency signals we observe are noncommensurate with apparent fine structure, those signals very likely represent nonradial oscillations of the white dwarf. The primary may belong to the well-known "ZZ Ceti" class of white dwarfs — pulsating DAs in a narrow temperature range (10800–12000 K, Mukadam et al. 2004). Several white dwarfs in CVs have been found with nonradial pulsations, and this is the commonly accepted interpretation. Members include GW Lib (van Zyl et al. 2004), SDSS 1610–01 (Woudt & Warner 2004), HS 2331+3905 (Araujo-Betancor et al. 2004), and possibly even WZ Sge (Robinson et al. 1978). It is not yet known whether this is exactly the ZZ Cet syndrome, or some close relative of it, or some new type of white-dwarf variability essentially triggered by accretion. These matters are discussed by Arras, Townsley, & Bildsten (2004).

In RE 1255 the DA absorptions and lack of flickering probably limit accretion light to less than 30% of the total in $V$. But the amplitude of the orbital signal (assuming we are correct in our interpretation) suggests that contribution to be at least 10%. So for definiteness we attribute 20±10% of $V$ light to accretion sources. This assigns $V$=19.2(2) to the white dwarf, and $V$=21(1) to accretion light. The latter could well be the small "infrared" component suggested by Figure 5; that would be plausible, since low-$\dot{M}$ disks cool mainly by optically thin emission which is rather flat with frequency (Tylenda 1981).

If the white dwarf is indeed a commonplace ZZ Cet star, with a normal temperature and mass (0.6 $M_\odot$), then it should have $M_v$=11.8(2). That would place RE 1255 at a distance of 300(80) pc. If the white dwarf is of 1 $M_\odot$, then it should have $M_v$=12.5(2), placing RE 1255 at a distance of 220(60) pc. The latter is probably the superior estimate, because the great breadth of the DA absorptions (Figure 2 of W96) suggests a high gravity.[4] However, both depend on the uncertain assumption that the white dwarf is of normal ZZ Cet temperature (11500 K); a proper formal analysis of the DA absorptions would yield a much improved estimate.

Because RE 1255 shows many close similarities to WZ Sge, a star of accurately known distance (43 pc: Thorstensen 2003, Harrison et al. 2003), a comparison may be useful for another distance estimate. The white-dwarf absorptions in WZ Sge yield $T$=15000 K (Sion et al. 1995), and spectroscopic and photometric evidence demand a massive white dwarf (1.0±0.2 $M_\odot$: Steeghs et al. 2001, Patterson et al. 2002). This implies $M_v$=12.2±0.6. But the visible "white dwarf" component has $V$=16.2±0.3; and the distance modulus of 3.2±0.1 then implies $M_v$=13.0±0.4. Assuming the white dwarf to be pure photospheric emission (not significantly enhanced by "accretion belts", for example), these estimates should be consistent. According them similar weight, we estimate $M_v$=12.6±0.5. Comparing to RE 1255, which probably also has a massive white dwarf of similar temperature, the observed $V$=19.2(2) then suggests a distance of 210(60) pc.

Another comparison star is GW Lib, also containing a pulsating DA in a short-period, low-$\dot{M}$ binary. GW Lib has the advantage that its visible light is dominated by the white dwarf, as is the case in RE 1255. At a trig-parallax distance of $104^{+30}_{-20}$ pc (Thorstensen 2003), GW Lib has $M_v$=13.0±0.5. If this is also true for RE 1255, then it lies at a distance of 160(50) pc.

These estimates — 220(60), 210(60), 160(50), and the much rougher 160(80) pc from proper motions — agree substantially, and we adopt a final figure of 180(50) pc.

This enables an estimate for the accretion light in RE 1255. We argued above for $V$=21(1), so this distance implies $M_v$=14.6±1.3 for accretion light. This signifies a remarkably low accretion rate. Since the star is evidently a dwarf nova, it presumably arises not from full disk accretion, but rather mass-transfer to the outer disk. So WZ Sge should again be a good comparison. The accretion light in that star has $V$=16.2(2) and hence $M_v$=12.9(3). Thus RE 1255 appears to be more intrinsically faint, by a factor of 3–6. Now these calculations do not account for differences in the angular dependence of emission, which are unknown and possibly large (because the binary inclinations are likely to be very different). Still, we expect the basic conclusion to be reliable, since the accretion light is much harder to see in RE 1255, competing with the light of probably rather similar white dwarfs.

---

[4] Though not necessarily as high as that deduced by W96 (log $g$>9). W96 based this on fitting line profiles to a white-dwarf model; but at the time, the star was ~1.8 mag above quiescence. An extra unwanted continuum source, if not subtracted, would drive the fit to much higher temperature and somewhat higher gravity — which it did (see Section 4).

The quiescent mass-transfer rate in WZ Sge is $\sim 1.0 \times 10^{15}$ g/s (Patterson 1984, Smak 1993, Patterson et al. 2002), and from that we expect $\sim 2.5 \times 10^{14}$ g/s for RE 1255. Now in the theory of CV evolution, $\dot{M}$ scales as $\dot{J}$; and if, as is thought likely for these short-period binaries, $\dot{J}$ is driven by gravitational radiation (GR), then $\dot{J}$ scales roughly as $q^2$. So $\dot{M}$ should scale as $q^2$, implying a $q$ about a factor 2 lower than that of WZ Sge. The latter appears to be $\sim 0.05 \pm 0.02$ (Patterson et al. 1998, Steeghs et al. 2001), leading us to expect $q$ near 0.025 in the case of RE 1255.[5] Let us see how that accords with available evidence from radial velocities.

## 4. SPECTROSCOPY AND BINARY STRUCTURE

### *4.1 THE OVERALL SPECTRUM*

W96 reported the following features in their spectroscopy:

(1) Broad H absorption, likely from a white dwarf photosphere, suggestive of high temperature and high gravity ($T_{\text{eff}} \sim 40000$ K, $\log g > 9$).

(2) Fairly narrow (FWHM 200 km/s) H emission, which they interpreted as arising from normal Doppler broadening in an accretion disk.

(3) Very stable velocities in the H emission. If this reflects the binary motion of the white dwarf, then $K_1 < 5$ km/s.

W96 interpreted (1) as signifying a massive and hot white dwarf, and (2) and (3) as signifying a very face-on binary. They assumed a 0.1 $M_\odot$ main-sequence secondary, and then needed a binary inclination less than 5º to obtain so stringent a limit on $K_1$.

This interpretation seems generally plausible to us, with several important qualifiers. *First*, the star was in 2003 1.0 mag (in *V*) fainter than the average value reported by W96, and 1.8 mag fainter than it was during their critical spectroscopy of January 1995, Thus their fluxes contain a significant contribution from accretion light — or at least the effect of "delayed light" from accretion energy, which white dwarfs in dwarf novae often exhibit (Long et al. 2004). Either case would corrupt the deduction of white-dwarf properties from the absorption lines. Simple subtraction of a continuum source from their spectra would have only a minor effect on $\log g$ (which is manifest mainly by line width), but would drastically lower the deduced $T_{\text{eff}}$ (which is manifest mainly by equivalent width). If for example we suppose that the contamination is 1.5 mag of continuum, then a proper subtraction would increase the white dwarf's Hγ equivalent width by a factor ~4, thereby lowering $T_{eff}$ (at log $g$=9) from 40000 to ~15000 K (see Table 1 of Koester et al. 1979). The photometric colors also establish that the white dwarf today is relatively cool. It may have even been (probably was) cool in 1995, but veiled by an extra ~1.5 mag of unwanted continuum.

---

[5] In this we assume that present-day "quiescence" is well-correlated with the long-term $\dot{M}$. This assumption can be invalidated by unrecognized cycles of nova outburst or secondary-star magnetism, and possibly other shortcomings of evolution theory.

　　*Secondly*, there is no particular reason to assume a 0.1 $M_\odot$ main-sequence secondary. After subtracting a 12000 K white dwarf from the flux distribution, scaled up by 1.0 mag to account for the apparent cooling between 1995 and 2003, the extra *JHK* light in Figure 5 is approximately flat with frequency — signifying a fairly hot source (~$10^4$ K). Above we argued that most of this is probably accretion light — optically thin emission from the disk and hot spot. However, in any spectral fit there is always room for a small cool component at the longest wavelength available; and we estimate that at most 30% of the residual *K* flux could come from a cool source. The secondary should then have $K$>18.8, or $M_K$>12.5 at the distance we favor. This is much too faint for a 0.10 $M_\odot$ main-sequence secondary, which has $M_K$=9.4 in the models of Baraffe et al. (1998). Indeed, it is marginally incompatible with the limiting case for H-burning at 0.075 $M_\odot$, for which the Baraffe et al. models yield $M_K$=11.4. Thus on luminosity grounds, the secondary should be of very low mass[6] — very likely below 0.075 $M_\odot$.

### 4.2  DEDUCTIONS FROM $K_1$

　　The emission-line radial velocities measured by W96 demonstrated that the lines hardly move; their Figure 6 suggests that $K_1$ does not exceed 5 km/s. Based on the discussion in Appendix A, we assume that this sets an upper limit of 5 km/s to the true value of $v_1 \sin i$.

　　A single-line spectroscopic binary satisfies

$$2\pi GM_1 (\sin i)^3 = P_o (K_1)^3 (1+q)^2/q^3, \qquad (1)$$

where $M_1$ is the white dwarf mass, $q=M_2/M_1$ is the mass ratio, and $i$ is the binary inclination. For RE 1255 this can be rewritten as

$$q^3/(1+q)^2 = 1.06\times10^{-6} / m_1 (\sin i)^3, \qquad (2)$$

where $m_1$ is in solar masses. All solutions have very low $q$, so we can approximate this as

$$q < 0.0102 / (m_1^{1/3} \sin i) \qquad (3)$$

This limiting $q$ varies with $m_1$ and $i$ as shown in Figure 6.

---

[6] Since the travails of CV evolution (heating by the primary, struggles with thermal equilibrium, nova eruptions) act only to *increase* the secondary's light beyond that allocated by nuclear physics and stellar structure. A good upper limit on light generally furnishes a pretty good upper limit on mass.

We have three other constraints. W96 required a high-mass white dwarf ($>1.2\ M_\odot$) to fit the Balmer lines. Above we have cited evidence disputing this conclusion. But even with today's much lower $T_{\text{eff}}$, the line profiles still look quite broad, suggesting a fairly high mass. Also, the upper limit on K light requires $m_2<0.075$, which we have expressed in Figure 6 as the curved line with arrows. What about sin $i$? If the observed emission-line breadth really arises from Keplerian motion in the disk, then the inclination must indeed be quite low. We can estimate how low by considering the emission lines of WZ Sge. The H$\alpha$ emission of the latter has FWZI~86 Å, whereas that of RE 1255 extends only to 15 Å. Adopting $i$~77º for WZ Sge, this suggests sin $i$ = 0.17 (or $i$=10º) in RE 1255. Of course we have no guarantee that the two stars resemble each other sufficiently to make this scaling, but the properties at quiescence do appear similar and it does produce a reasonable result. For the curves in Figure 6, this suggests $q<0.06$.

The apparent existence of an orbital modulation supplies yet another rough constraint on $i$. Face-on CVs should have no orbital signal at all, and low-$\dot{M}$ CVs at average inclinations (~40–65º) usually manage to concoct an orbital signal of ~0.15–0.35 mag full amplitude (after correcting for the white-dwarf contribution). In RE 1255 the signal is of 0.028 mag full amplitude; but only 20% of the light seems to come from accretion, so this corrects to 0.14 mag full amplitude. It is hard to see how such a signal can emerge from a binary with $i$ as low as 5º — or even 10º. With a higher $i$, Figure 6 requires a $q$ considerably lower than 0.06. That's consistent with the value of 0.025 suggested from a luminosity argument at the end of Section 3.

## 5. EVOLUTIONARY STATUS: WHENCE AND WHITHER?

### 5.1 OTHER STARS OF VERY LOW q

This star may have the lowest $q$ and lowest quiescent $\dot{M}$ of any dwarf nova yet studied. Yet $P_{\text{orb}}$ appears to be well-separated from the period minimum, where most very faint dwarf novae reside. Does it have any colleagues in this respect?

#### 5.1.1 GD 552 = NSV 25966

Yes, probably it does. A very likely colleague is GD 552, a blue emission-line star of high proper motion. Greenstein & Giclas (1978) estimated a distance of 70 pc, which implies $M_V$~12.2. This suggests a very low $\dot{M}$. Hessman & Hopp (1990, hereafter HH) found a radial-velocity period of 0.07134(12) d, with $K_1$=16±4 km/s. Yet GD 552 possesses the usual signatures of high-inclination binaries: doubled and broad (±2000 km/s, HH) lines, and an orbital modulation in the photometry (unpublished CBA data). High $i$ and low $K_1$ can only mean a very small $q$. Let's try to make this constraint more quantitative.

For GD 552 Eq. (1) yields

$$q/(1+q)^{2/3} = 0.00204\ K_1/(m_1^{1/3} \sin i), \qquad (4)$$

where $K_1$ is in km/s. Solutions with $K_1$=16 km/s are shown as the curved lines in the upper frame of Figure 7. The inclination is bounded above by the lack of eclipses ($i<70°$) and below by the line profiles (roughly $i>30°$). Based on Appendix A, the curved lines should more properly be considered upper limits on $q$. HH measured the separation of the doubled peaks to be ±440(40) km/s from line center. If we knew exactly how to interpret these peaks, we could use this as an additional constraint (mainly on $M_1$ and $i$).

In the simplest model of emission lines in quiescent CVs, the half-separation is assumed to be $v_d \sin i$, where $v_d$ is the Keplerian velocity at the outer edge of the disk When confronted with real data, this assumption encounters difficulties (Smak 2001, Horne & Marsh 1986, HH). We try here to bypass these difficulties — without needing to understand or solve them — by calibrating them out with measurements from binaries with well-determined parameters. There are seven stars suitable for this calibration: eclipsing CVs with sufficiently strong constraints on $M_1$, $M_2$, and the half-separation of the double peaks. This data is presented in Table 3. We assume that the latter is $v_d \sin i$, where $v_d = (GM_1/R_{em})^{1/2}$ and $R_{em}$ is the emission-line radius of the disk. The emission-line radius should fit inside the Roche radius of the primary, which we approximate as

$$R_{roche} = 0.436\, a\, q^{-0.12}. \qquad (5)$$

This is a simple power-law approximation to the "volume equivalent" radius of the Roche lobe (Kopal 1959), and is accurate to 2% in the relevant domain of $q(0.02 \rightarrow 0.20)$. We write

$$R_{em} = k\, R_{roche} \qquad (6)$$

and solve for $k$. A little algebra yields

$$k = (932/v_d)^2\, m_1\, m^{-1/3}\, P_{hr}^{-2/3}\, q^{0.12}, \qquad (7)$$

where $v_d$ ($=v_{peak}/\sin i$) is in km/s, $m$ and $m_1$ are in solar masses, and $P_{hr}$ is the orbital period in hours. Table 3 collects the published constraints on all the quantities on the right side of the equation, and the last column shows the resultant estimate for $k$.

These values for $k$ average 0.81±0.08. Now the precise meaning of this is unclear. For example, the observed peak velocities vary systematically over the Balmer series, with the higher members showing higher velocity. These differences are large, and mostly not understood (though see Horne & Marsh 1986 for a lucid discussion). So it is not correct to say "this is the radius of the outer disk" — but rather, this is *the effective radius that reproduces the separation of Hβ emission-line peaks in the simple model for the best-studied stars*. We will use this estimate in interpreting the profiles in other double-peaked stars.

Now we invert the problem and solve (7) for GD 552 with $k=0.7–0.9$ and $v_d=440±40$ km/s/sin $i$. This yields

$$\sin i = (0.44 \rightarrow 0.57)\, m^{1/6}\, m_1^{-1/2}\, q^{-0.06}. \qquad (8)$$

Since $q$ can only be in the range $0.03 \rightarrow 0.07$ and the dependence on $q$ is very weak, we can approximate $q$~0.05 and simplify this to

$$\sin i = (0.52 \rightarrow 0.70)\, m_1^{-\frac{1}{3}}, \tag{9}$$

and then solve for $i(m_1)$ as shown in the lower frame of Figure 7. The lower (and upper) envelope in this frame sets the upper (and lower) bound of allowed solutions. This implies that $q$ is in the range 0.040–0.055 if the "true" $K_1$ is 16, or lower (since $q$ scales as $K_1$) if the true $K_1$ is lower.

HH favored a different solution. Their analysis of $q$ is somewhat similar to ours, yielding $q$=0.06–0.09. But by requiring the secondary to be a 0.15 $M_\odot$ ZAMS star, they were then forced to invoke a 1.4 $M_\odot$ white dwarf, and then to tilt the binary nearly face-on ($i$=16º) to reduce the Doppler shifts enough to match observation. This is essentially a solution at the extreme upper right of the top frame of Figure 7. It seems implausible to us, since the line profiles (width and doubling) are really a major credential *forbidding* a low inclination. It also seems impossible to match the required low luminosity with their $M_2$. A 0.15 $M_\odot$ ZAMS star should have $M_K$=8.3 (Baraffe et al. 1998), or $K$=12.6 at a distance of 70 pc. Yet the 2MASS survey gives $K$=14.5, and this must be an extreme upper limit to the secondary's $K$ brightness, since the measured $K$ image contains also light from accretion, the white dwarf, and an unwanted foreground/background star. For the secondary itself, we regard $K$>16 as a conservative upper limit, and $K$>17 as very likely. These imply $M_K$>11.7 and 12.7 — fainter than the brightness expected at the Kumar limit (11.4). We express this luminosity constraint by the $m_2$<0.075 arrowed curve in Figure 7. Thus the totality of evidence much prefers low $m_2$, low $q$, and a moderate (not low) inclination. And indeed, that solution also appears as the best choice in HH's Figure 10; they were just unwilling to embrace it, perhaps because of reluctance to accept a substellar secondary.

These estimates would be greatly improved by a parallax measurement for GD 552. Unfortunately, the star's high proper motion ran it in front of a field star ten years ago, so this will have to wait a while. Nevertheless, the high proper motion certainly suggests a very nearby star. Using available sky survey images and our own recent astrometry, we measured

$$\mu_x = +124(4) \text{ mas/yr}$$
$$\mu_y = -41(4) \text{ mas/yr},$$

which implies a median distance of 50 pc for CVs in the Galactic disk (using the model of Thorstensen 2003). This is certainly consistent with the 70 pc previously estimated by Greenstein & Giclas.

### 5.1.2  1RXS J105010.3–140431

Another candidate is 1RXS J105010.3–140431, which shows a spectrum much like that of RE 1255 — a DA white dwarf with emission lines. The white dwarf absorption wings are nearly stationary, and the very low $K_1$ therefore implies a very low $q$ (Mennickent et al. 2001, hereafter M01). How low?

M01 reported an upper limit of 20 km/s to the motion of the Hα emission-line wings, and a detection of motion with $K_1=4\pm1$ km/s in the absorption lines of the white dwarf. In principle, the latter is just the desired measurement, $v_1 \sin i$ without contamination by disk motions. But the cited evidence in their Figure 5 does not include a power spectrum or periodogram, and therefore does not actually establish the significance of the detection. Instead we will use the upper limit from the emission-line wings.[7]

Solving Eq. (1) for $P_o=0.0615$ d and $K_1=20$ km/s, we obtain

$$q = 0.038/(m_1^{1/3} \sin i), \qquad (10)$$

again in the low-$q$ limit which the observations require. These solutions are shown in the upper frame of Figure 8. The binary inclination must be fairly high, because the emission lines are broad and doubled, with a half-separation of 540±40 km/s. Repeating the above analysis with $k=0.7–0.9$, we obtain

$$\sin i = (0.52 \rightarrow 0.66)\, m^{1/6}\, m_1^{-1/2}\, q^{-0.06}, \qquad (11)$$

or

$$\sin i = (0.62 \rightarrow 0.79)\, m_1^{-1/3} \qquad (12)$$

in the ballpark of reasonable solutions ($q\sim0.05$). With equation (10) this implies $q=0.048–0.061$. And it's an upper limit, so we conclude $q<0.06$.

Astrometry of this star is worth a mention. Combining our recent images with available sky surveys, we measured

$$\mu_x = -187(4) \text{ mas/yr}$$
$$\mu_y = -8(4) \text{ mas/yr},$$

which implies a nearby star if it belongs to the Galactic-disk CV population. We express this with a preliminary distance estimate of 100(50) pc.

The $K$ magnitude is 15.7±0.3 in 2MASS, and we estimate from the *UBVRIJH* fluxes that the sources of blue light (white dwarf + accretion) contribute at least half of the $K$ light. Thus we estimate that the secondary has $K>16.4$, or $M_K>11.4$ for a distance of 100 pc. This upper limit is just the $K$ light expected at the Kumar limit.

### 5.1.3 DWARF NOVAE

---

[7] Their Figure 5 does suggest a very low upper limit, however. This is definitely worth a follow-up study!

Four dwarf novae appear to have good credentials as period-bouncers, based on the small fractional period excesses in their superhumps. The idea is that the observed period excess $\varepsilon$ signifies the size of the secondary star's perturbation on the disk, and is roughly proportional to $q$ (P01, and discussed below).[8] Since $\varepsilon$ can often be measured to ~10%, $q$ can be inferred to similar accuracy if the calibration is good enough. Those four stars are listed in Table 4, with relevant details and sources. The three stars previously discussed are included in this gallery of good period-bouncer candidates.

Table 4 contains one star not yet adequately reported in the literature — ASAS 153616–0839.1 (ASAS 1536), a dwarf nova with a first recorded eruption in 2004. Its credentials must be considered weaker since there are still no published studies. But its apparent low $\varepsilon$ (from unpublished CBA data) is a strong claim to membership, so we tentatively accept it. Our astrometry yields

$$\mu_x = +48(4) \text{ mas/yr}$$
$$\mu_y = -72(4) \text{ mas/yr.}$$

For a disk CV, the corresponding probability distribution peaks at 60 pc, though remains still fairly high beyond 100 pc. The standard candle of the dwarf-nova eruption itself suggests a distance more like 150 pc. We compromise on an estimate of 120(50) pc.

### 5.2 BINARY EVOLUTION: $q(P_{orb})$

Secondary stars in CVs lose mass monotonically as they age, so the progression of $M_2$ should reflect the arrow of evolution. But there is no general way to measure $M_2$, except for a handful of eclipsing stars. The mass ratio $q$ is a possible surrogate, because white-dwarf masses in CVs seem to average ~0.8 $M_\odot$ for all the classes that have been studied (Webbink 1990, Smith & Dhillon 1998). We have previously cited evidence that the rms dispersion in $M_1$ does not exceed ~25% (P03), so $q$ is likely an acceptable surrogate. Unfortunately, opportunities to measure $q$ directly are also rare; there are only a few eclipsers, only a few detections of the secondary, and the radial velocity of the primary is tainted by the issues discussed in Appendix A. An interesting surrogate for $q$ is the fractional period excess $\varepsilon$ of superhumps. This is measurable for ~90 stars to good precision, does not depend on inclination, and in principle actually measures the gravitational force exerted by the secondary (through the tidal force rather than the usual monopole term). This is a promising route to $q$.

It does, however, need to be calibrated. We used

$$\varepsilon = 0.18\, q + 0.29\, q^2, \tag{13}$$

an improvement on the earlier calibration at large $q$ (P05). We then converted $\varepsilon$ to $q$ in the current list of superhumping stars, added the $q$ constraints from our seven stars, and produced the $q(P_{orb})$ relation seen in Figure 9. The solid curve in Figure 9 is that of theory, assuming mass transfer powered strictly by GR, and secondaries which start on the main sequence and are gradually driven out of thermal equilibrium by the ongoing mass loss (Kolb & Baraffe 1999).

---

[8] The present paper is in a series of studies of CV evolution: Patterson 1998, 2001; Patterson et al. 2003, 2005 - which we abbreviate as P98, P01, P03, P05.

Figure 9 suggests a fairly simple pattern of short-period evolution, in which binaries start anywhere (e.g. at the top) and evolve towards minimum period in a manner much like that predicted by theory. They reach minimum and bounce to longer periods. Near minimum, bounce credentials become obscure, because the curve is locally vertical. But the seven stars appearing to be on the lower branch are the best candidates: AL Com, WZ Sge, EG Cnc, RX 1050, ASAS 1536, GD 552, and RE 1255, in order of increasing apparent age.[9]

Although the points in Figure 9 do not track the theory well, they do suggest a curve *similar* to that of theory. The theory curve is slightly too high, and reaches a period minimum slightly too short. Previous discussions (P98, P01, P03) have shown that a minor adjustment to the secondary's mass-radius relation remedies both discrepancies, and the dashed curve shows the predicted path with the P01 mass-radius relation. In fact, such an adjustment is mandatory, because all the well-determined mass–radius pairs show that the secondaries are too big (by 10–30%, see Figure 4 of P01) to be theoretical main-sequence stars. Barker & Kolb (2003) calculate how such an adjustment might result from an additional angular momentum sink (beyond GR).

*5.3 SUMMARY*

Figure 9 offers evidence that, as suggested by theory, there really is a lower branch to the $q(P_{\mathrm{orb}})$ curve — *period bouncers*. The candidacy of these seven stars is further strengthened by other columns in Table 4. In particular, the accretion light in these stars is very, very faint; the absolute magnitudes in Table 4 are essentially the faintest known among CVs. This is another expected feature of period-bouncers, since we expect $\dot{M} \propto q^2$. Finally, the distances in Table 4 are noteworthy. Since these stars are intrinsically faint and very rare erupters, selection effects discriminate heavily against their discovery. Nevertheless, this collection includes the nearest known CV (WZ Sge), and possibly the second nearest (GD 552). It seems likely that this is a very, very populous class.

**6. THE EUV ERUPTION**

Another remarkable feature of RE 1255 is the great EUV eruption of 1994. Dahlem et al. (1995) report a peak 112–200 Å flux of $7 \times 10^{-9}$ ergs/cm$^2$/s — exceeding that of any persistent source in the EUV sky, even the mighty HZ 43. From nondetection in a deep EUVE pointing a year later, Drake et al. (1998) estimated that the EUV flux brightened by a factor exceeding 50000 in the 1994 eruption. At 200 pc, this implies a peak EUV luminosity exceeding $4 \times 10^{34}$ erg/s — probably exceeding SS Cygni, the most prolific soft X-ray (SX) emitter among dwarf novae, although this is sensitive to unknown details of the spectral shape (Ponman et al. 1995,

---

[9] Not necessarily *real* age, of course. A binary should begin its CV life at essentially an arbitrary point on the evolution curve, depending on the accidents of prior nuclear evolution and angular momentum loss in the common-envelope phase. "Apparent age" adopts the fiction that all stars begin mass transfer at the same point. The term "period bouncer" adopts the same fiction — whereas in reality, some CVs may be born on the lower branch.

Mauche et al. 1995).

Only three other dwarf novae have appeared as very bright EUV transients: SS Cygni, U Geminorum, and VW Hydri. Why this strange quartet — the three best-studied dwarf novae in the sky, and a complete newcomer? The reason is probably the one discussed by Patterson and Raymond (1985, hereafter PR): that strong EUV detections should be limited to stars which are nearby or on very transparent sight-lines (since $F_{euv}$ is very sensitive to interstellar absorption) and stars with massive white dwarfs (since $L_{euv}$ scales approximately as $M_1^{5.5}$). SS Cyg, U Gem, and VW Hyi are all nearby dwarf novae. The first two possess white dwarfs of certifiably high mass, and VW Hyi lies on a very transparent sight-line (Polidan et al. 1990). RE 1255+266 probably has a high $M_1$, evinced by the broad lines. And its galactic latitude of 89° places it square in the midst of the "Local Chimney", a region of exceptionally low $N_H$ (Welsh et al. 1999). A transparent sight-line means that the penalty of distance is only $d^{-2}$, a much milder penalty than is usually exacted near the galactic plane.

So that much seems plausible and simple. This particular star just happened to erupt when the WFC was observing and scanning that particular strip of the sky. Not every dwarf nova could make this much EUV flux, but every one satisfying these two conditions (mass and sight-line) could. The origin of the luminosity would presumably lie in the boundary layer (BL) where the accretion disk grazes the white dwarf surface. Roughly half of the total accretion luminosity should be released in that tiny belt around the white dwarf's equator.

Perhaps. This is probably the main story of the RE 1255 eruption. But it was not merely the first erupting dwarf nova discovered in this way by the WFC (which had a short lifetime) — it is the only one discovered in the past thirty years by *any* SX/EUV telescope. It might need yet another distinguishing credential to explain this oddity.

That credential might come from its unusually low binary inclination, discussed above. Low $i$ may well be favored for emergent BL emission, basically because of circumstellar absorption. This is obvious near the orbital plane, since the accretion disk is very opaque. But in practice, obscuration by structures associated with the disk may well extend quite high above the plane. Light curves of U Gem (Mason et al. 1988, Naylor & la Dous 1997) and WZ Sge (Patterson et al. 1998) show that X-ray opacity effects of gaseous structures associated with the hot spot extend at least 25° above the disk, and also quite far in azimuth from the point of stream impact. At this height, the inferred $N_H$ is still as high as $(3-100) \times 10^{20}$ cm$^{-2}$ — plenty high enough to block soft X-rays. And those observations refer to quiescence. It does not strain credulity to imagine that disks in a high-$\dot{M}$ state might well concoct opacity sources which effectively block soft X-rays even quite far from the orbital plane.

## 7. BOUNDARY LAYERS IN DWARF NOVAE

In 1985, PR compared the theory of optically thick boundary layers to observations of high-$\dot{M}$ CVs, and found satisfactory agreement. How has the situation changed in the last twenty years?

### *7.1 NEW DEVELOPMENTS IN THEORY*

There have been several changes on the theoretical side. PR used simple power-law relations to estimate the effects of $M_1$, $N_H$, and $T_{BL}$ on observable soft X-ray flux; but the BL theory of Popham & Narayan (1995, hereafter PN) solved for the size of the emitting region — the most critical parameter in comparing with data — by explicitly accounting for boundary conditions and temperature gradients in the BL. Both studies reached the same basic conclusions (compare Figure 9 and Table 2 of PN to Figures 2–4 of PR): that strong soft X-ray detection of the BL requires low $N_H$ and high $T_{BL}$; that the latter requires an exceptionally high $M_1$ and $\dot{M}$; and that possibly only SS Cyg and U Gem actually satisfy these requirements. PN also examine the role of white-dwarf rotation. The predicted SX emission is reduced, of course, if the white dwarf is rapidly rotating; and it is also reduced during the time that the white dwarf (or its "accretion belt") is spinning up, since energy is required for the spin-up. PN calculated about a factor of 3 reduction in $L_{BL}$ for a rotation frequency assumed to be half the Keplerian rate.

PR and PN both assumed a blackbody spectral shape in estimating the observable SX emission. Figure 3 of PR shows the basic problem in producing soft X-rays: a BL at $2 \times 10^5$ K emits only 10% of its luminosity in soft X-rays, and the observable component falls to ~1% even for quite modest absorption ($10^{20}$ cm$^{-2}$). This is very unpromising. But in reality, the BL is more like a hot high-gravity stellar atmosphere with solar abundances. How does this change the expected emission? Barman et al. (2000) present models of this type. The results are somewhat complex, but the main change is to create absorption edges on the Wien tail, and these edges always reshuffle the flux back to lower energies (Figures 1 and 2 of Barman et al. illustrate this). And since in this case the expected $T_{BL}$ regime (100–300 kK) places the practical observing window ($\lambda < 100$ Å) on the Wien tail, the blackbody assumption tends to overestimate the observable soft X-ray flux. So: the blackbodies are too cool, the stellar atmospheres are worse, and even a modest extinction can make these boundary layers invisible to our X-ray telescopes.

### 7.2 NEW OBSERVATIONS

PR summarized the state of observations c. 1984, near the end of the Einstein Observatory era. Since then, there have been four telescopes with comparable or better sensitivity to soft X-rays: ROSAT, EUVE, and the two current telescopes with their low-energy gratings (XMM and Chandra). There have been several summaries of the ROSAT data (Richman 1996, van Teeseling & Verbunt 1994); they mainly confirm the Einstein results: strong SX emission from SS Cyg and U Gem, and upper limits from all other relevant targets (high-$\dot{M}$ disk accretors). EUVE has been more productive on bright sources with low $N_H$; results are summarized by Mauche (2002). Spectroscopy of U Gem in outburst was fit by Long et al. (1996) with $T_{bb}$=110–140 kK, and this implies $L_{BL} \sim 4 \times 10^{34}$ erg/s, or $L_{BL}/L_{disk} \sim 1$. Similar observations of SS Cyg gave $T_{bb}$=230–350 kK, implying $L_{BL}/L_{disk} \sim 0.07$ (Mauche, Raymond & Mattei 1995). In principle these observations give far superior measures of $T_{bb}$, therefore $L_{BL}$, than anything else available. However, as pointed out by Mauche & Raymond (2000), the detailed spectra are very complicated and heavily punctuated by emission and/or absorption; so the significance of the blackbody fits is not as clear as one would like.

Strangely enough, the fainter sources may turn out to be more informative. Spectroscopy of OY Car and WZ Sge has been obtained in superoutburst, and the observed continuum flux is much too low to be consistent with a simple $L_{BL}=L_{disk}$ description (Mauche & Raymond 2000, Wheatley & Mauche 2004). However, the reason is almost certainly the high orbital inclination of these eclipsing binaries. The SX flux of both stars is dominated by emission lines, suggesting origin in a wind; apparently there is no clear line of sight to the white dwarf, and we see only photons scattered into the line of sight by the wind (Mauche & Raymond 2000). In such a case the observations do not clearly and directly constrain the issue of a luminous BL component. However, Mauche & Raymond modelled the scattering and found a best fit to the OY Car spectrum with $T_{bb}$=90–130 kK, implying $L_{BL}=(1-4)\times10^{34}$ erg/s and $L_{BL}/L_{disk}\sim1$.

### *7.3 SUMMARY: BOUNDARY LAYERS REVISITED*

Despite persistent references to "missing" boundary layers, the evidence does not support this — and never did.[10] Bright disk-accreting CVs are usually undetected in soft X-rays, for the two basic reasons cited by PR and PN: the temperatures are too low, and the absorption is too high. This is evinced by the extensive HEAO–1 and HEAO–2 surveys — and by the "continuum" temperatures deduced from EUVE spectroscopy, which appear to be ~100–160 kK for all stars except SS Cyg. Even for blackbodies, this is too cool to leak significant flux into a typical soft X-ray bandpass; and solar-abundance model atmospheres at such temperatures seem to be worse in that respect.

Actual measurement of $L_{BL}$ in these apparently cool sources is not feasible, however, since the estimated $T_{BL}$s deposit the vast majority of flux into the EUV (say >200 Å), where interstellar and circumstellar absorption would hide it. Further progress along these lines may come from modelling the emission lines, as done by Mauche & Raymond (2000) for OY Car.

### 8. SUMMARY AND THE VIEW AHEAD

1. We report time-series photometry of the optical counterpart of RE 1255+266. At least three signals noncommensurate in frequency are present (with periods of 1344, 1236, and 668 s), and probably others. The (full) amplitudes range from 0.015 to 0.06 mag. These signals are likely due to nonradial pulsation of the white dwarf.

2. There is also a weak signal (0.028 mag full amplitude) at a period of 0.08298(12) d, which we interpret as the orbital period of the underlying binary.

3. The *UBVRI* colors in 2003 demonstrate that the white-dwarf temperature was then near 12000 K, where ZZ Ceti stars live. Allowing for uncertainty in log *g*, we estimate $T_{eff}$=13000±2000 K. The star was 1.0–1.8 mag brighter in early 1995, either because the white dwarf had not sufficiently cooled from outburst, or because of extra accretion light (or, probably, both). This softens W96's log *g*>9 constraint; but the gravity should still be fairly high, to explain the broad wings of the absorption lines.

---

[10] In the sense of implying some crisis for the energetics. Some shortfalls in the apparent $L_{BL}$ are obviously possible, and the complex spectral shapes are a puzzle deserving much study.

4. Several methods of estimation are consistent with a distance of 180±50 pc. This implies that accretion light shines with a measly $M_v$=14.6±1.3, which suggests a very low mass-transfer rate of $\sim 3 \times 10^{14}$ g/s. The secondary is unseen at all wavelengths, which sets a limit of $M_K$>12.5. This is marginally too faint (by 1 mag) to be compatible with the Kumar limit at 0.075 $M_\odot$.

5. RE 1255 seems likely to be a dwarf-nova of very long recurrence period, as suggested by W96 and W00. We assume this, as well as the interpretation that the narrowness and immobility of the H emission lines signify a low binary inclination. W96's observed $K_1$<5 km/s limit then probably implies $q$<0.06. This is supported by our study of published $K_1$ estimates in CVs generally, which is presented in Appendix A and suggests that accretion disks generally act to produce an observed $K_1$ which is an upper limit to the true value of $v_1 \sin i$.

6. Several other clues, not as firm, favor an even lower $q$. If driven by GR, the very low $\dot{M}$ suggests $q \cong 0.025$. And the apparent existence of an orbital signal in the light curve suggests $i > \sim 20°$, which then implies $q < \sim 0.03$.

7. At this $P_{orb}$, such a $q$ implies that the binary is a candidate "period-bouncer". Other stars possessing good credentials are GD 552 and RX 1050, as well as four known dwarf novae. These stars seem to have all the attributes expected of period-bouncers, and probably reflect a very high population of such objects. Identifying these stars did not require a latest-greatest survey; a few weeks in the library, a little photometry, and a good pencil sharpener did the job.

8. We discuss the EUV eruption of 1994 in the context of a dwarf-nova model. At the likely distance of ~180 pc, the peak observed EUV luminosity is $\sim 3 \times 10^{34}$ erg/s. This can be understood as the emission from a disk-star boundary layer accreting at $\sim 10^{-8}$ $M_\odot$/yr, assuming $T$~300 kK. It could therefore be a good "poster child" for theories of the boundary layer. We discuss other EUV and soft X-ray observations of high-$\dot{M}$ disk-accreting white dwarfs, and conclude that the majority of evidence does not support the idea that there is substantial "missing energy" from the boundary layers. Most such allegations arise from comparison of real data with inappropriate choices of boundary-layer temperature (too high) and interstellar/circumstellar absorption (too low). With a very low binary inclination, a very happy position at the Galactic Pole, and possibly a high-mass white dwarf, RE 1255+266 appears be ideally equipped for an intense and observable soft X-ray outburst every few decades.

9. Period bouncers are a new arrival in the CV zoo. The cage was prepared long ago (Paczynski & Sienkiewicz 1983); and theorists have been observed pacing back and forth, wondering why observers have not managed to bring back specimens for the cage. Several recent papers (P01, Littlefair et al. 2003) discuss the difficulties in certifying such stars. The most common search strategy (identifying spectral features of the secondary) has yielded none, and has the demerit of being incalculable, since period-bounce secondaries are allotted luminosity only from minor processes (reflection, departures from thermal equilibrium) likely to be quirks of individual binaries. Search methods based on *mass* are better, and here we suggest two (superhumps and radial velocities) that have yielded seven good candidates. The $q(P_{\rm orb})$ distribution is likely to be a very useful way to acquire more specimens for the zoo.

J.P. thanks Chris Mauche and Ed Jenkins for helpful discussions, Eve Armstrong and Maddy Reed for assistance with observations, and grants from the NSF (AST–00–98254) and NASA (HST–GO–09406.01A and NNG04GA83G) for funding this research. J.R.T. thanks the NSF for support through AST–99–87334 and AST–03–07413. This research has made use of the USNOFS Image and Catalogue Archive operated by the U.S. Naval Observatory, Flagstaff Station.

# APPENDIX A
# CORRUPTIONS OF $v_1 \sin i$

It is by now well-known (e.g., Stover et al. 1981, Thorstensen et al. 1991) that radial velocities of CV emission lines are untrustworthy indicators of the true dynamical motions of the individual stars. The reasons are not yet known in detail, but are basically related to confusion by the large velocities in the disk. The emission lines arise from the disk, the disk is inevitably somewhat asymmetric (due to perturbation by the secondary, and due to the "hot spot" where the mass-transfer stream strikes the outer disk), and that asymmetry moves around with $P_{orb}$. And since the Doppler motions in the disk and hot spot are much larger than the expected motion of the disk center (white dwarf), attempts to measure the latter are easily distracted by the former. In short-period systems, the measured values of $K_1$ — which purports to be the true $v_1 \sin i$ — are especially prone to corruption, because the expected values of $v_1 \sin i$ are quite small. In WZ Sge, for instance, the measured values of $K_1$ are in the range 37–50 km/s (summarized by Patterson et al. 1998), whereas we know from photometric evidence and from the spectroscopic detection of the secondary's lines that the true $v_1 \sin i$ is very likely in the range 20–25 km/s (e.g. Steeghs et al. 2001).

Nevertheless, it is a reasonable hypothesis, consistent with evidence to date, that these spurious values of $K_1$ always yield an overestimate (i.e. overestimate $v_1 \sin i$). We show the evidence for this in Table A1 and Figure A1, which contain all short-period CVs with an observed value of $K_1$ and a reasonably well-constrained mass ratio. For the "observed" $K_1$ we generally use published estimates based on measuring the wings of emission lines, a technique developed by Schneider & Young (1980) and later refined by Shafter (1983a, b) and Marsh (1988). This is regarded as the best general technique to avoid the very great distortions afflicting the centers of emission lines. We omit published $K_1$ estimates with very large uncertainty, and those based on "whole line" measurements.

The most accurate values for $q$ and $i$ come from photometric constraints on eclipsing CVs; these are indicated in Table A1 by "E", and the stars are denoted by name in Figure A1 since the error bars tend to be small. The great majority of CVs do not eclipse, but many show "superhumps" in their light curves. It is easy to measure the superhump periods with good accuracy, and the excess of that period over $P_{orb}$ appears to be well-correlated with $q$ — and therefore is a good surrogate for $q$ (P01, P05). We use the superhump measurements to infer $q$, and then Kepler's Third Law to yield a predicted value for $v_1$:

$$v_1 = 572 \text{ km/s } (m_{0.8})^{1/3} q(1+q)^{-2/3} (P_{hr})^{-1/3}$$

where $m_{0.8}$ is the mass of the primary in units of 0.8 $M_\odot$. Two other ingredients are needed. We assume a standard value of $m_{0.8}=1$; this is consistent with previously published average values (e.g. Smith & Dhillon 1998), and the weak dependence on mass immunizes us from serious error. Inclination is a much more serious worry. Edge-on binaries ($i>75°$) announce their inclination with great fanfare, since they eclipse; for these we use published values, or a default $i=80°$. Noneclipsing CVs may still reveal a moderately high inclination, through less emphatic means: their emission lines may be doubled or exceptionally broad, or their light curves may show a large orbital modulation. For this class we assign broad limits, typically $i=40–70°$ but varying slightly depending on individual details (amplitude of orbital signal, extent of line doubling, line breadth). Finally, there is the "low inclination" class; some are sure to be of genuinely low $i$, while others simply lack suitable relevant data, and still others may just be squeamish about revealing their inclination. For this class we merely assign a conservative upper limit, typically $i<60°$. The predicted value of $K_1$, shown as the abscissa in Figure A1, is then $v_1 \sin i$.

The uncertainties are dominated by our ignorance of $i$ — except for eclipsing systems, where $i$ is fairly well known and essentially unimportant (since $\sin i \sim 1$ for $i>75°$). The many upper limits (all arising from uncertainty in $i$) make this figure slightly hard to read, but the moral for our purposes is clear enough: all stars with small error bars fall to the right of the ideal $v_1 \sin i = K_1$ line, and all other stars are consistent with that location — subject to the uncertainty in $i$.

Some improvements to Table A1 and Figure A1 could be made. To minimize clutter we do not show (although we tabulate) individual errors in $K_1$. We also do not regard the constraints on $i$ as firm for each star, since the tabulated data span a huge range in quality (and since the relevant physics of line formation is poorly known). But this coarse analysis suffices to demonstrate our main point: that the disk's corruption of $v_1 \sin i$ generally, and perhaps universally, acts to increase its value.

For the interested reader, good discussions of this corruption have been given by Hessman et al. (1989) and Marsh et al. (1987).

TABLE 1
LOG OF OBSERVATIONS

| Run | Date | JD (2450000+) | Duration (nights/hrs) | Telescope | Filters |
|---|---|---|---|---|---|
| 1 | 2000 May | 1686–9 | 4/13 | 1.3 m | clear |
| 2 | 2001 December | 2270–1 | 2/4 | 2.4 m | clear |
| 3 | 2002 April | 2378–86 | 7/42 | 1.3 m | clear |
| 4 | 2002 June | 2450–2 | 3/5 | 2.4 m | clear |
| 5 | 2004 April | 3101–10 | 5/19 | 1.3 m, 2.4 m | clear, $V$, $R$ |
| 6 | 2004 May | 3135 | 1/5 | 1.3 m | clear |
| 7 | 2004 June | 3163–6 | 4/13 | 1.3 m | clear |

CALIBRATED MAGNITUDES: $V$=19.00(3), $U-B$=–0.7, $B-V$=0.11, $V-I$=0.24(7).

TABLE 2
SIGNIFICANT PERIODICITIES FOUND

| Run | Date (JD 2450000+) | Frequencies (cycles/d) | Frequency Error (cycles/day) |
|---|---|---|---|
| A. Individual Nights ||||
| 1 | 1686 | 64.5  129.4  136.4  262.0  328.5  723.2 | 2.0 |
| 1 | 1687 | 62.5  129.4  273.0 | 2.0 |
| 2 | 2270 | 69.3  139.5 | 2.4 |
| 2 | 2271 | 72.1  138.2 | 2.4 |
| 3 | 2378 | 62.3  130.0  193.4 | 1.0 |
| 3 | 2379 | 128.4(fat) | 1.0 |
| 3 | 2381 | 124.0 | 1.8 |
| 3 | 2382 | 64.5  129.3  221.8 | 0.7 |
| 3 | 2383 | 65.0  129.5 | 0.8 |
| 3 | 2385 | 127.9 | 1.0 |
| 3 | 2386 | 65.7  130.9 | 0.9 |
| 4 | 2451 | 122.0  716.0 | 2.6 |
| 6 | 3135 | 30.1 | 1.0 |
| B. Merged Nights ||||
| 1 | 1686–89 | 22.74/23.73  63.27/64.26/62.28  128.94/129.92  136.10/137.07/135.11  326.78/325.79/327.78 | 0.08 |
| 2 | 2270–71 |  | 0.11 |
| 3 | 2378–86 | 12.07  64.27  129.28  maybe others, see text | 0.02 |
| 7 | 3163–66 | 28.83/29.82 | 0.08 |

TABLE 3
BASIC DATA ON ECLIPSERS

| Star | $P_{orb}$ (d) | $q$ | $M_2$ ($M_\odot$) | $M_1$ ($M_\odot$) | $v_{peak}$ (km/s) | $i$ (°) | $k$ | Source |
|---|---|---|---|---|---|---|---|---|
| WZ Sge | 0.05669 | 0.045(15) | 0.05 | 1.1 | 720 | 77 | 0.92 | Steeghs et al. 2001, Krzeminski & Kraft 1964 |
| OY Car | 0.06312 | 0.10(1) | 0.08 | 0.80 | 770 | 83 | 0.69 | Wood et al. 1989, Schoembs & Hartmann 1983 |
| V2051 Oph | 0.0624 | 0.19(3) | 0.15 | 0.78 | 765 | 83 | 0.73 | Baptista et al. 1998 |
| HT Cas | 0.07365 | 0.15(1) | 0.09 | 0.61 | 600 | 81 | 0.88 | Horne et al. 1991, Wood et al. 1992 |
| IY UMa | 0.07391 | 0.125(8) | 0.10 | 0.79 | 700 | 84 | 0.76 | Patterson et al. 2000, Steeghs et al. 2003 |
| Z Cha | 0.07450 | 0.145(15) | 0.09 | 0.65 | 660 | 82 | 0.77 | Wood & Horne 1990 |
| U Gem | 0.17691 | 0.36(2) | 0.41 | 1.12 | 540 | 69 | 0.85 | Long & Gilliland 1999, Smak 2001 |

NOTE: By convention, this $v_{peak}$ is the half-separation of the Hβ emission. The Hα separation is typically 17% lower, and Hγ is 12% greater.

TABLE 4
PERIOD–BOUNCER CANDIDATES

| Star | $P_{\rm orb}$ (d) | $d$ (pc) | $m_v, M_v$ (quiescent) | quiescent $M_v$ (accretion) | $\log <\dot{M}>$ (g/s) | $T_{\rm rec}$ (yr) | $q$ | Sources |
|---|---|---|---|---|---|---|---|---|
| AL Com | 0.05667(1) | 400(120) | 20.2, 12.1 | 13.0 | 15.0 | 20 | 0.064(16) | Patterson et al. 1996, Szkody et al. 2003 |
| WZ Sge | 0.05669(<1) | 43(3) | 15.4, 12.1 | 12.9 | 15.0 | 30 | 0.05(1) | Patterson et al. 2002 |
| EG Cnc | 0.05997(12) | 350(90) | 18.8, 11.1 | ~12 | 15.4 | 20 | 0.035(15) | Patterson et al. 1998, Szkody et al. 2002 |
| RX 1050 | 0.0615(1) | 100(50) | 17.6, 12.6 | 13.5 | 14.7 | >5* | <0.06 | M01, this paper |
| ASAS 1536 | 0.06414(2) | 120(50) | 18.0, 12.5 | 13 | 15 | >5* | 0.035(20) | this paper |
| GD 552 | 0.07134(6) | 70(30) | 16.6, 12.3 | 12.5 | 15.2 | >10* | <0.055 | HH, this paper |
| RE 1255 | 0.08285(12) | 180(50) | 19.0, 12.4 | 14.6 | 14.5 | >5* | <0.06 | this paper |

* Lower limit based on nondetection of two outbursts, or in some cases one — and on estimating the frequency of monitoring by variable-star observers.

TABLE A1
ESTIMATED $v_1 \sin i$ VERSUS MEASURED $K_1$

| Star | $q$ | $v_1$ (km/s) | $i$ (°) | $K_1$ (km/s) | Constraints | Sources |
|---|---|---|---|---|---|---|
| V844 Her | 0.114(7) | 55(6) | 45–70 | 97(11) | SH, LD | Thorstensen et al. 2002 |
| WZ Sge | 0.045(10) | 26(5) | 77 | 42(7) | E, SH, LD, S | Steeghs et al. 2001, Patterson et al. 1998 |
| SW UMa | 0.115(8) | 55(6) | 30–60 | 47(8) | SH, PQ, S | Semeniuk et al. 1997, Nogami et al. 1998, Shafter et al. 1986 |
| WX Cet | 0.097(6) | 46(5) | 30–60 | 57(4) | SH, S, PQ | P03, Thorstensen et al. 1996, Kato et al. 2001 |
| T Leo | 0.111(6) | 52(5) | 30–60 | 135(8) | SH, S, PQ | Lemm et al. 1993, Shafter & Szkody 1984 |
| KV Dra | 0.110(7) | 52(7) | 45–70 | 54(9) | SH, LD, S | P03, Nogami et al. 2000 |
| AQ Eri | 0.131(7) | 60(7) | 40–65 | 37(4) | SH, S | Thorstensen et al. 1996, Kato 1991 |
| CP Pup | 0.083(8) | 43(5) | 40–70 | 115(8) | SH, S | Patterson & Warner 1998, O'Donoghue et al. 1989 |
| V1159 Ori | 0.145(5) | 63(4) | <55 | 40(6) | SH, S, PQ | Patterson et al. 1995, Thorstensen et al. 1997 |
| V2051 Oph | 0.129(7) | 59(6) | 70–80 | 111(12) | E, SH, S | P03, Baptista et al. 1998, Watts et al. 1986 |
| V436 Cen | 0.101(5) | 47(6) | 50–70 | 59(8) | SH, LD, S | Gilliland 1982, Semeniuk 1980 |
| VY Aqr | 0.098(4) | 45(4) | 40–65 | 49(4) | SH, S, PQ | Patterson et al. 1993, Thorstensen & Taylor 1997 |
| HO Del | 0.130(9) | 59(7) | <50 | 54(15) | SH, S | P03, Kato et al. 2003 |
| BC UMa | 0.137(4) | 64(4) | 45–70 | 56(6) | SH, LD, PQ, S | P03 |
| OY Car | 0.098(2) | 45(3) | 81 | 61(8) | E, SH, PQ, S | Hessman et al. 1989, 1992 |
| EK TrA | 0.145(6) | 67(5) | 30–60 | 61(7) | SH, S | Mennickent & Arenas 1998, Vogt & Semeniuk 1980 |
| ER UMa | 0.142(5) | 65(4) | <55 | 48(4) | SH, S, PQ | Thorstensen et al. 1997, Kato & Kunjaya 1995 |
| UV Per | 0.111(4) | 51(5) | <45 | 29(4) | SH, S, PQ | P98, Thorstensen & Taylor 1997, Udalski & Pych 1992 |
| AK Cnc | 0.163(6) | 73(6) | 25–55 | 50(7) | SH, S | Mennickent et al. 1996, Arenas & Mennickent 1998 |
| DM Lyr | 0.129(7) | 59(6) | <60 | 37(5) | SH, S | Thorstensen & Fenton 2003, Nogami et al. 2003 |
| SX LMi | 0.155(6) | 70(5) | 50–70 | 57(11) | SH, S | Wagner et al. 1998, Nogami et al. 1997 |
| SS UMi | 0.160(5) | 71(4) | 40–65 | 44(4) | SH, S | Thorstensen et al. 1996, Chen et al. 1991 |
| KS UMa | 0.113(4) | 51(7) | <50 | 47(5) | SH, S | P03 |
| RZ Sge | 0.140(5) | 62(5) | 40–60 | 62(5) | SH, LD, S | P03, Bond et al. 1982 |
| TY Psc | 0.155(5) | 69(4) | 40–70 | 53(5) | SH, PQ, S | P98, Thorstensen et al. 1996 |
| IR Gem | 0.156(14) | 70(12) | <50 | 30(6) | SH, S, PQ | Szkody, Shafter & Cowley 1984 |
| CY UMa | 0.161(7) | 71(5) | 30–60 | 65(4) | SH, S | Harvey & Patterson 1995, Thorstensen et al. 1996 |
| FO And | 0.156(12) | 69(8) | <60 | 43(4) | SH, S | Thorstensen et al. 1996, Kato 1995 |
| VZ Pyx | 0.147(5) | 65(5) | 25–55 | 53(5) | SH, S | Thorstensen 1997, Kato & Nogami 1997 |
| CC Cnc | 0.204(10) | 86(7) | 40–60 | 60(6) | SH, S | Thorstensen 1997, Kato & Nogami 1997 |
| OU Vir | 0.151(4) | 75(5) | 79 | 103(20) | E, PQ, SH, S | Mason et al. 2002, Feline et al. 2004a |
| HT Cas | 0.148(5) | 65(5) | 81 | 115(6) | E, S, SH | Zhang et al. 1986, Marsh et al. 1990, Young et al. 1981 |
| VW Hyi | 0.148(3) | 64(3) | 50–70 | 78(14) | SH, PQ, S | Schoembs & Vogt 1981, van Amerongen et al. 1987 |
| Z Cha | 0.161(4) | 70(3) | 82 | 88(8) | E, PQ, SH, S | Wade & Horne 1988, Marsh et al. 1987, Warner & O'Donoghue 1988 |
| QW Ser | 0.148(7) | 64(7) | 45–70 | 85(8) | SH, LD, S, PQ | P03 |
| WX Hyi | 0.155(4) | 67(4) | 50–70 | 67(6) | SH, PQ, S | Bailey 1979, Schoembs & Vogt 1981 |
| BK Lyn | 0.201(9) | 85(4) | 30–60 | 100(20) | SH, S, PQ | Skillman & Patterson 1993, Ringwald et al. 1996 |
| RZ Leo | 0.155(4) | 66(4) | 50–70 | 129(35) | SH, PQ, LD | P03, Ishioka et al. 2001 |
| SU UMa | 0.145(4) | 62(3) | 30–60 | 58(4) | SH, S, PQ | Udalski 1990, Thorstensen et al. 1986 |
| HS Vir | 0.195(9) | 82(6) | 40–60 | 59(12) | SH, S, PQ | P03 |
| V503 Cyg | 0.184(9) | 77(4) | <60 | 71(10) | SH, S, PQ | Harvey et al. 1995 |
| CU Vel | 0.146(7) | 57(6) | <60 | 49(10) | SH, S | Mennickent & Diaz 1996, Kato 2003 |
| TY PsA | 0.180(7) | 74(5) | 50–70 | 68(8) | SH, S, LD | Barwig et al. 1982, O'Donoghue & Soltynski 1992 |
| DV UMa | 0.153(4) | 64(6) | 84 | 140(10) | E, PQ, S, SH | Patterson et al. 2000, Feline et al. 2004b |
| YZ Cnc | 0.226(5) | 89(5) | <60 | 41(20) | SH, S, PQ | Shafter & Hessman 1988, Patterson 1979 |
| GX Cas | 0.191(6) | 77(5) | 45–70 | 96(18) | SH, LD, S | P03 |
| TU Crt | 0.174(7) | 71(5) | 45–70 | 85(9) | SH, LD, S | Mennickent et al. 1998, P03 |
| V348 Pup | 0.253(8) | 95(6) | 80 | 100(6) | E, SH, S | Rolfe et al. 1998, Baptista et al. 1996, Rodriguez-Gil et al. 2001 |
| V795 Her | 0.290(10) | 104(5) | <50 | 70(20) | SH, S, PQ | Shafter et al. 1990, Patterson & Skillman 1994 |
| V592 Cas | 0.248(10) | 89(5) | <50 | 52(4) | SH, S | Taylor et al. 1998 |
| TU Men | 0.276(8) | 97(6) | 45–70 | 90(15) | SH, S | Stolz & Schoembs 1984, Mennickent 1995 |
| AH Men | 0.325(10) | 108(6) | 30–60 | 140(25) | SH, S, PQ | Buckley et al. 1993, Patterson 1995 |
| TT Ari | 0.312(10) | 103(5) | <40 | 54(10) | SH, S, PQ | Skillman et al. 1998, Thorstensen et al. 1985 |
| V603 Aql | 0.222(7) | 79(5) | <40 | 22(8) | SH, S | Patterson et al. 1997 |
| PX And | 0.328(10) | 106(5) | 65–80 | 162(25) | E, SH, S | Thorstensen et al. 1991 |
| UU Aqr | 0.272(10) | 85(5) | 70–85 | 121(7) | E, SH, S | Patterson et al. 2005, Baptista et al. 1994, Diaz & Steiner 1991 |

NOTE. — Explanation for constraints: SH = superhumps (can provide an accurate $q$), E = eclipses (usually provides accurate $q$ and $i$), PQ = photometry at quiescence (suggests a rough constraint on $i$), LD = line-doubling (usually requires a fairly large $i$), S = spectroscopy (line-breadth, S-wave amplitude constraining $i$). Most $q$ estimates come from applying Eq. (13) to the measured ε and its error. The error in Eq. (13) itself is unknown, and thus not included. It matters little in this context, since the total error is nearly always dominated by ignorance of $i$.

## FIGURE CAPTIONS

FIGURE 1. — *Upper frames*: two selected light curves, condensed to 88 s/point. *Lower frames*: nightly power spectra of nine nights with clear detection of periodic signals. Nights are labelled with truncated JD (true JD – 2452000).

FIGURE 2. — Average power spectrum of the seven nights of Run 3, with detected signals labelled by frequency (±0.7). In order of increasing frequency, we refer to these as $\omega_0$, $\omega_1$, and $\omega_2$.

FIGURE 3. — Two-night power spectrum of Run 2. Aliasing is much too severe to permit unambiguous choice of frequency in the three regions. The various aliased peaks are separated by 1.03 c/d, and their frequencies are measured to ±0.1 c/d. The arrows point to the only choices fitting a 1:2:3 scheme. It is also possible, of course, that the signals are not commensurate.

FIGURE 4. — Seven-night power spectrum of Run 3. *Upper frame*: close-up of $\omega_0$, the signal we propose as $\omega_{orb}$. *Lower two frames*: close-ups of $\omega_1$ and $\omega_2$, the apparent white-dwarf pulsations. The power-spectrum window is inset. At least in the vicinity of $\omega_1$, there appears to be power in the data not present in the window — evidence of a nearby (weaker) signal.

FIGURE 5. — Squares represent the flux distribution of RE 1255. The *UBVI* magnitudes were obtained in January 2003, and the *JHK* magnitudes were obtained in April 1995 (10 months after outburst). The continuous curve shows the flux distribution (on the scale at right) of G 226–29, a pulsating white dwarf with $T_{eff}$=12300 K, log $g$=8.3. The *JHK* fluxes appear to show a moderately blue ($F_\nu \sim \nu^0$) excess over a white dwarf fit; but the size of that excess is probably smaller than shown here, since the star was ~1 mag brighter in April 1995 (see Table 4 of W96).

FIGURE 6. — The curves with inclination labels show solutions for RE 1255 with $K_1$=5 km/s. Acceptable solutions must lie below the $m_2$=0.075 curve (arrows), and should also have a white dwarf of fairly high mass (say $m_1$>0.9). These considerations roughly establish $q$<0.06. Other arguments, discussed in the text, suggest $i$~10° and $i$>10°.

FIGURE 7. — *Upper frame*, $q(m_1)$ solutions for GD 552 with $K_1$=16 km/s. Acceptable solutions must lie below the $m_2$=0.075 curve (arrows). *Lower frame*, the envelope of allowed solutions that can reproduce the doubled emission lines. All solutions satisfying both requirements have $q$<0.06. The combined constraints require $q$=0.048(7), but this is an upper limit since the measured $K_1$ may be too large.

FIGURE 8. — *Upper frame*, $q(m_1)$ solutions for RX 1050 with $K_1$=20 km/s. Since the true $K_1$ may be lower, these are actually upper limits on $q$ (and $q \propto K_1$). *Lower frame*, the envelope of allowed solutions that can reproduce the doubled emission lines. The combined constraints require $q$=0.055(7), but this is an upper limit.

FIGURE 9. — Dependence of "measured" $q$ on $P_{orb}$, for $P_{orb}$<3.5 hours. We use $\varepsilon$ as a surrogate

for $q$ [Eq.(13) and P05]. The low dispersion in the upper branch suggests rms errors in $q$ of <20% (see discussion in Sec 6.2 of P03). The period-bouncer candidates are denoted by name. Upper limits come from radial-velocity studies (detection or upper limit on $K_1$), and the square symbol is an estimate based on $\dot{M}$. Superposed is a (solid) theoretical evolution curve, assuming mass transfer driven by gravitational radiation and "main-sequence" secondaries (Kolb & Baraffe 1999, adjusted to a mean $M_1=0.8\ M_\odot$). The dashed curve is the corresponding evolution with a more realistic mass-radius relation for CV secondaries (P01). The evidence for period bounce lies in the existence of the lower branch.

FIGURE A1. — Comparison of the observed $K_1$ in radial-velocity studies with the $v_1 \sin i$ predicted for the orbital motion of the white dwarf. The data are drawn from Table A1. All the points with small errors are eclipsers, identified by name. The open circles indicate *upper limits* to the predicted $v_1 \sin i$; these are stars to which we can only assign an upper limit to $i$. In a perfect world, all the stars would lie along the diagonal line.

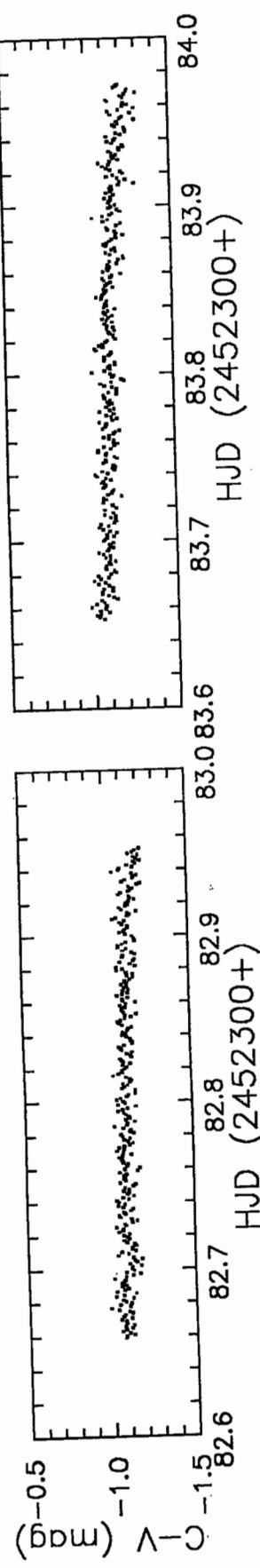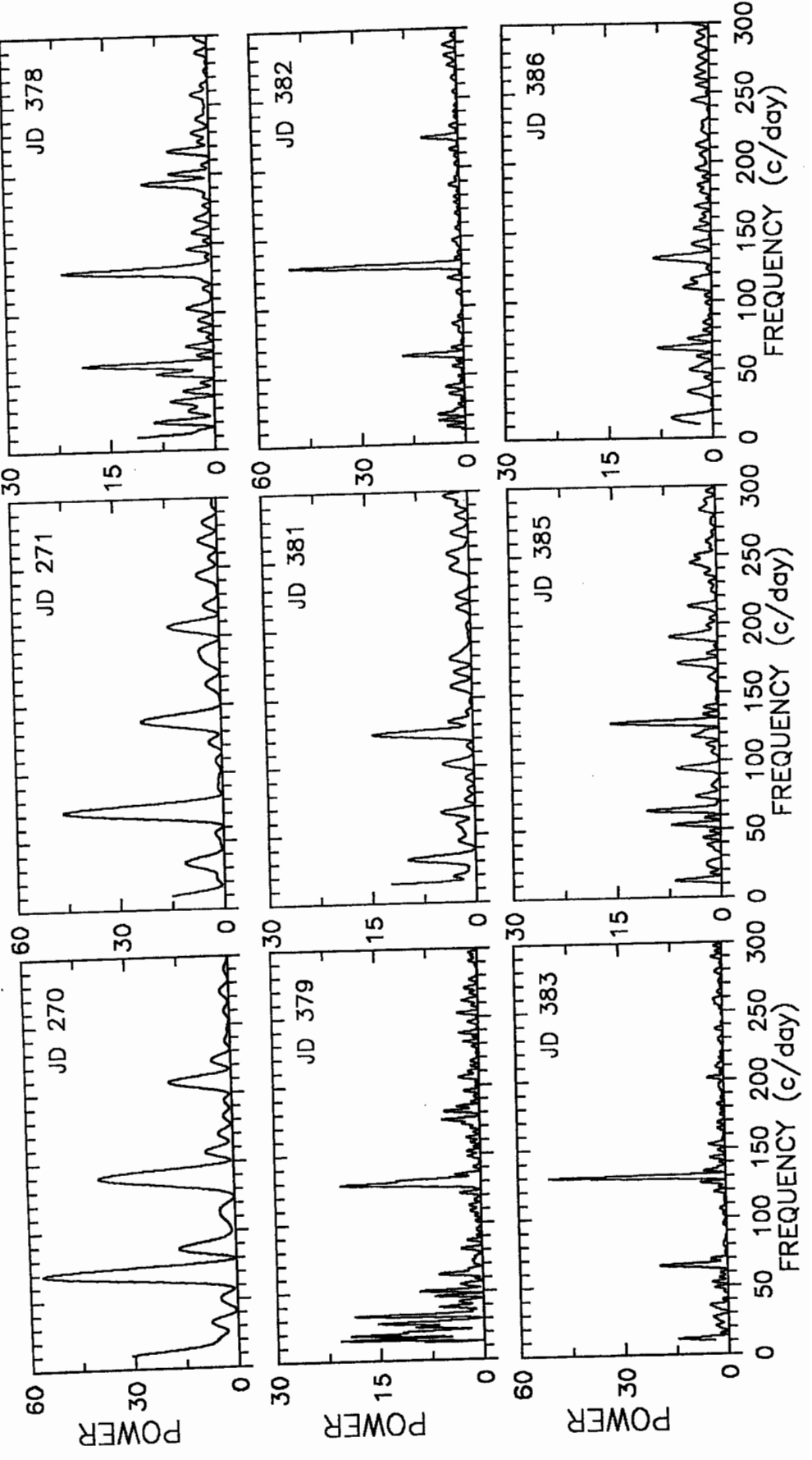

Fig 1

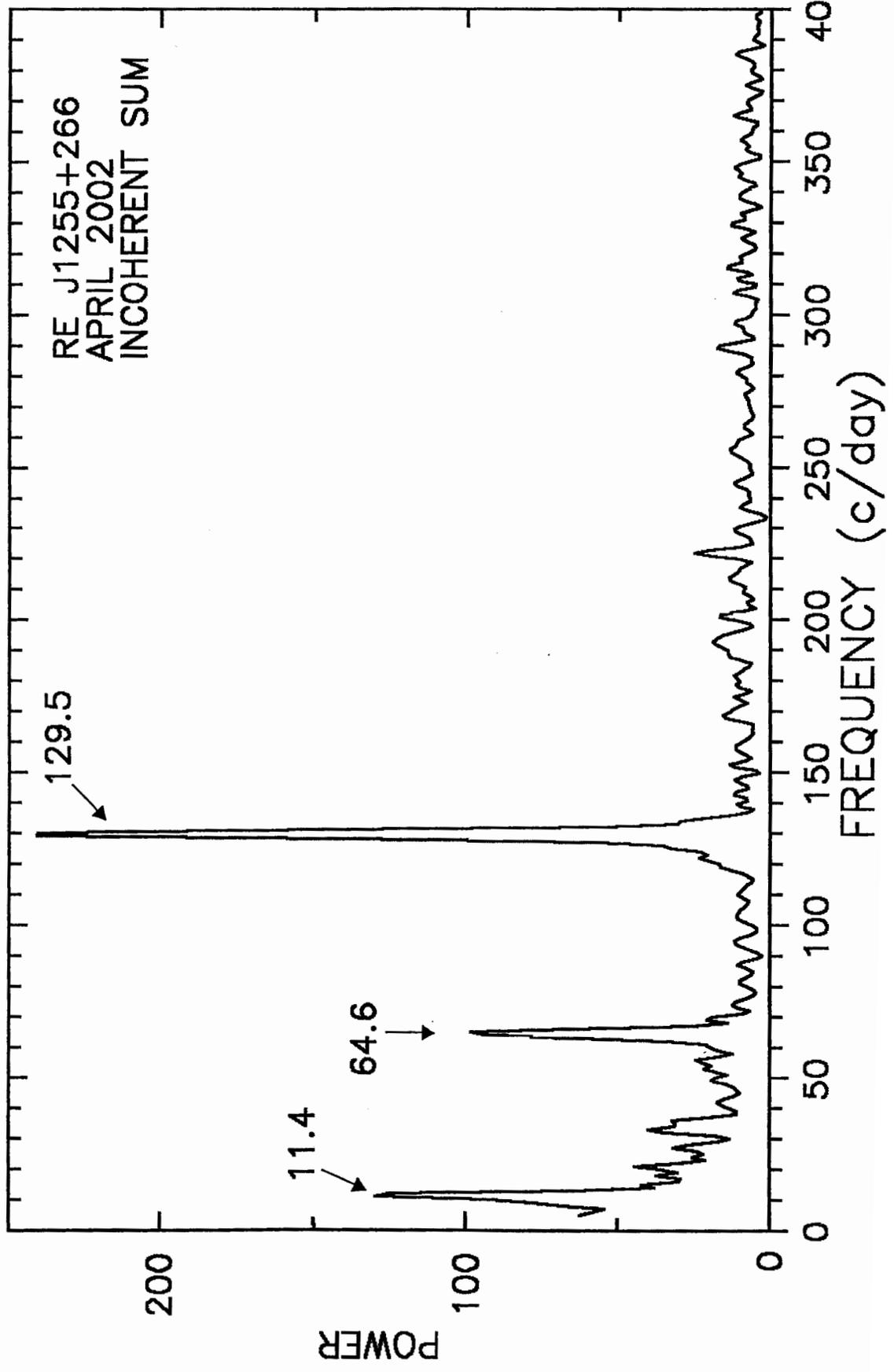

Fig 2

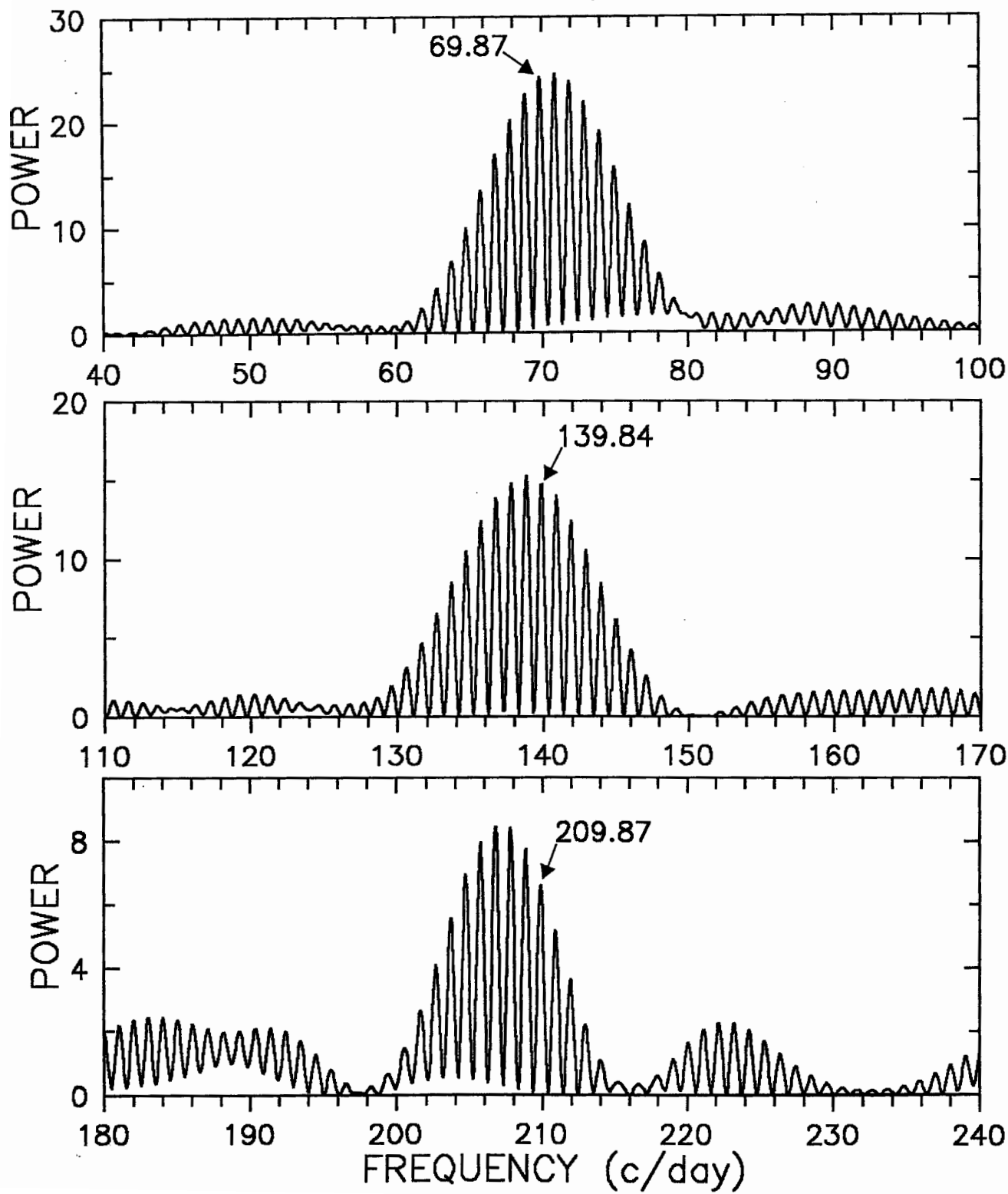

Fig 3

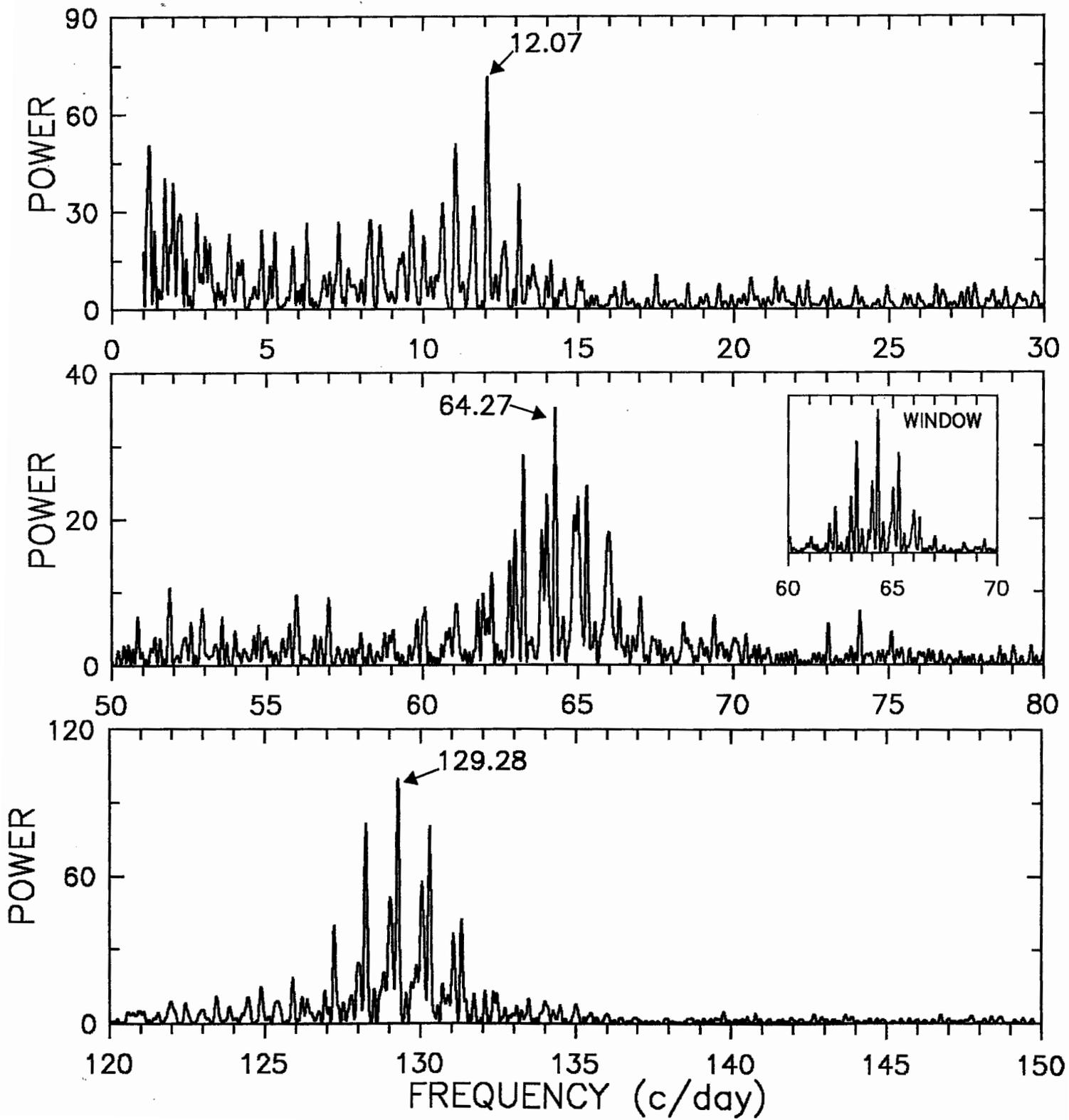

Fig 4

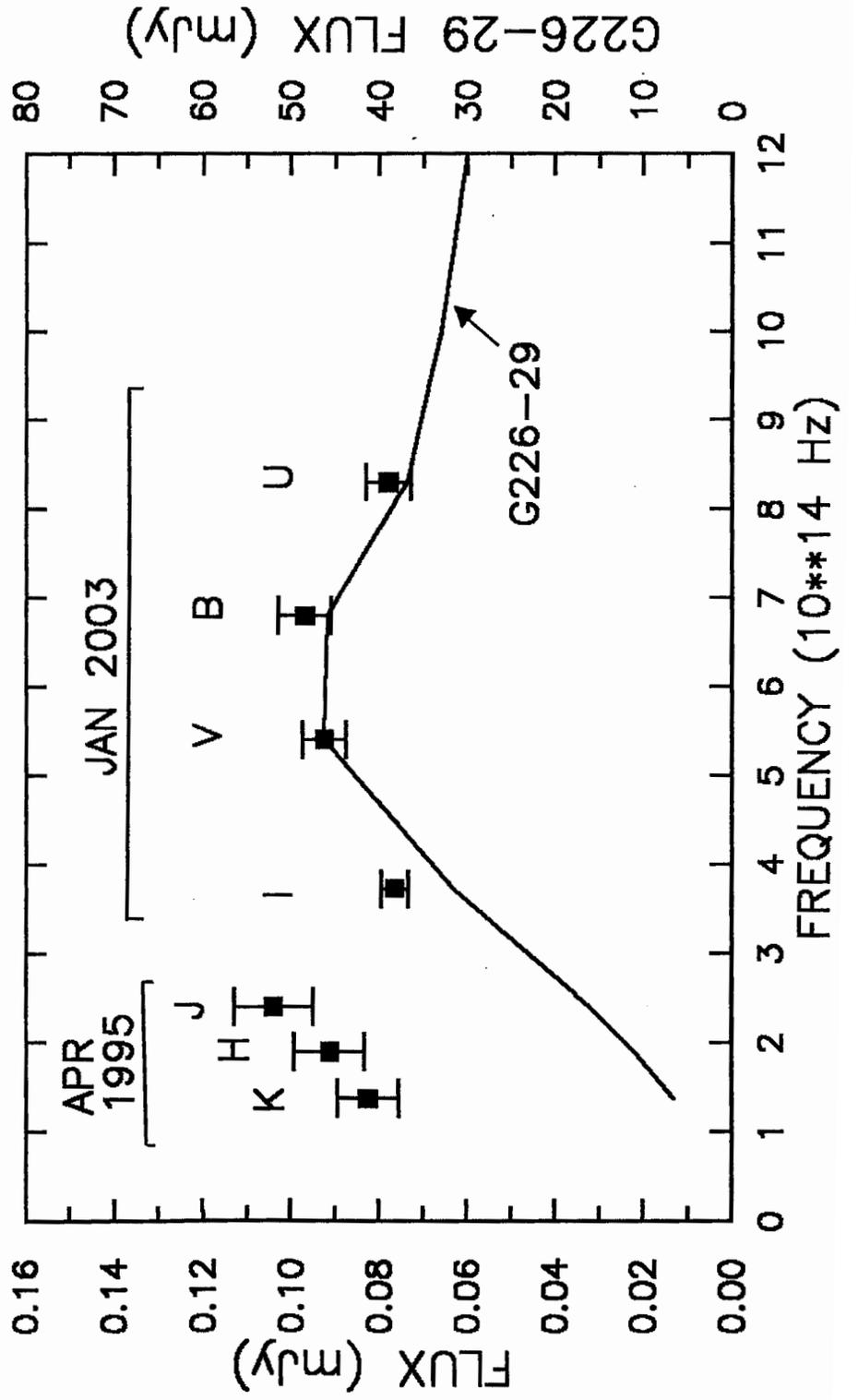

Fig 5

Fig 6

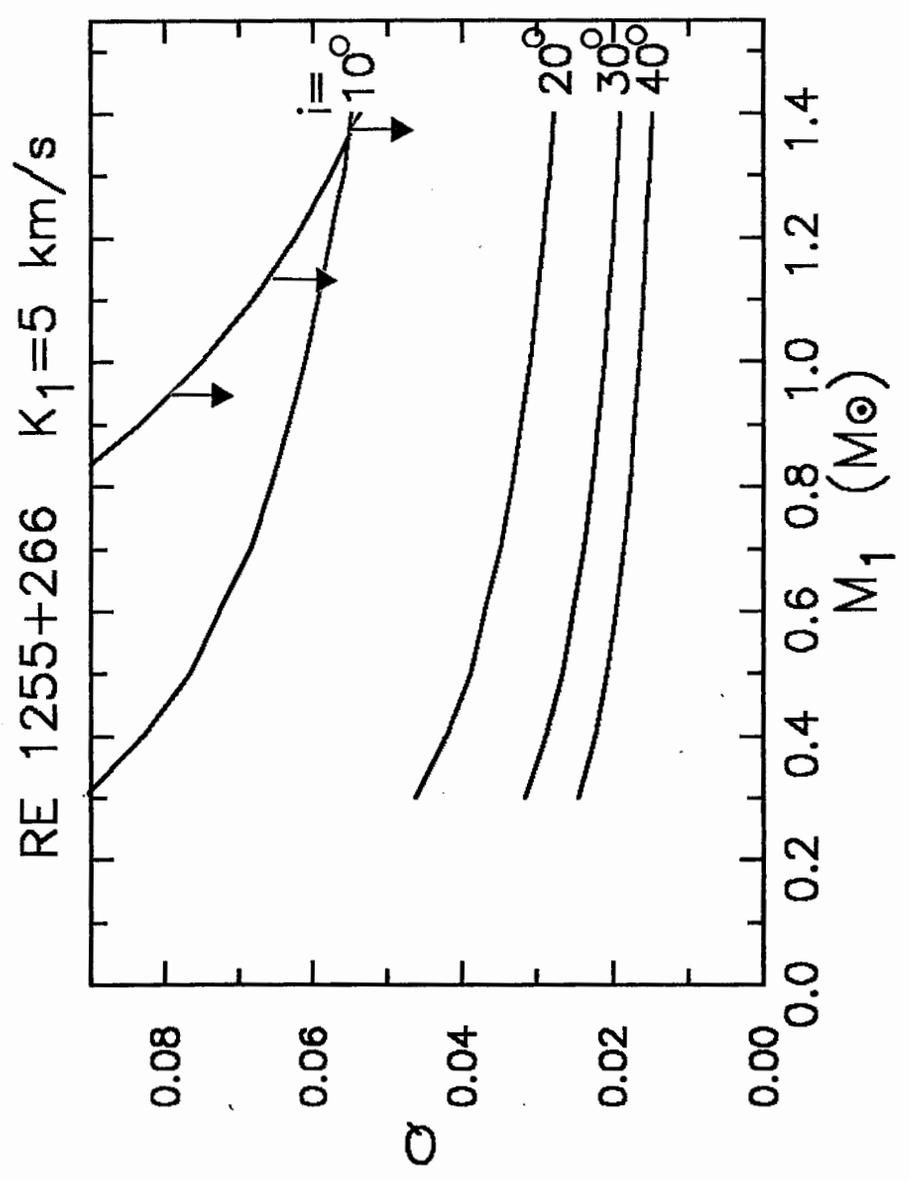

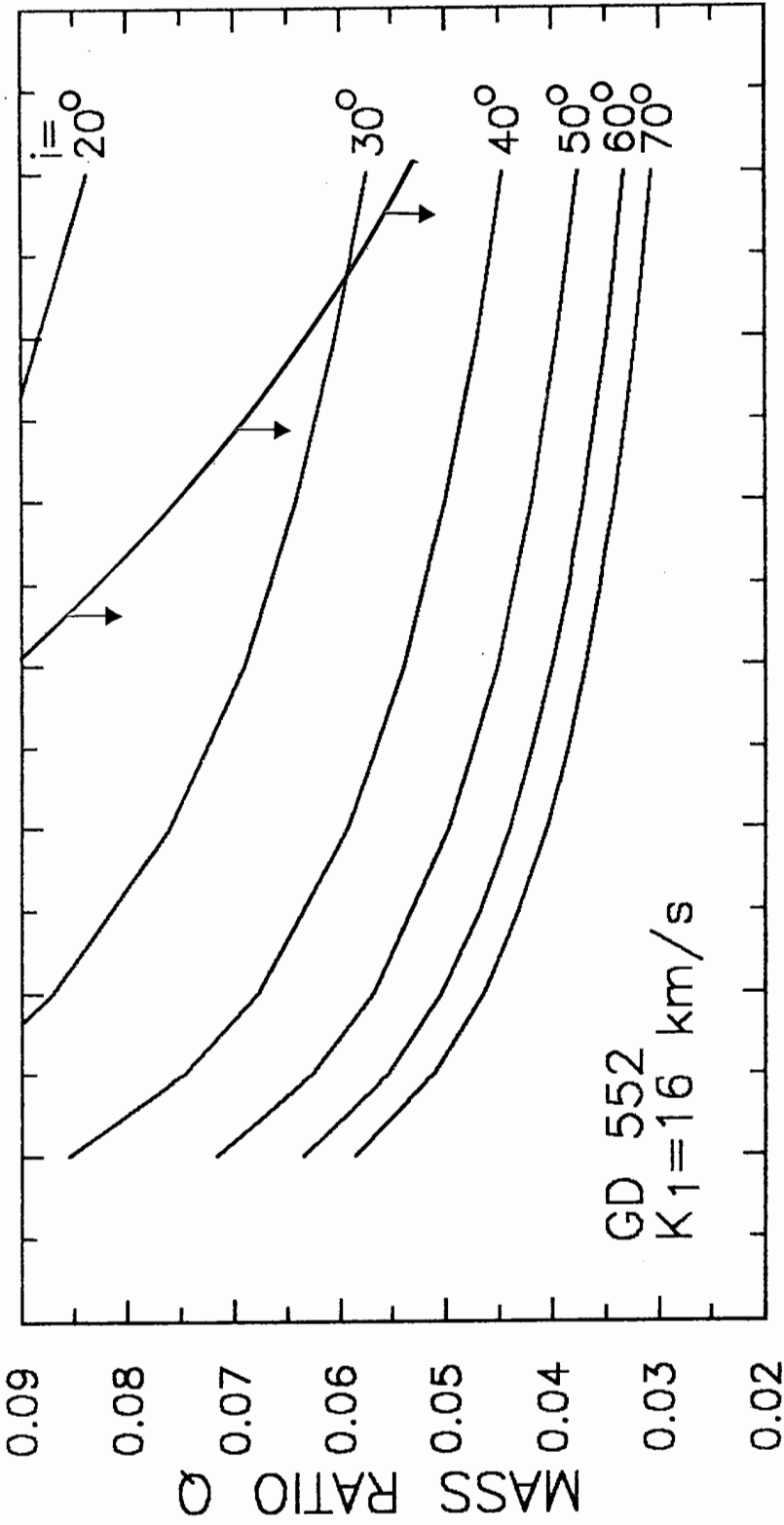
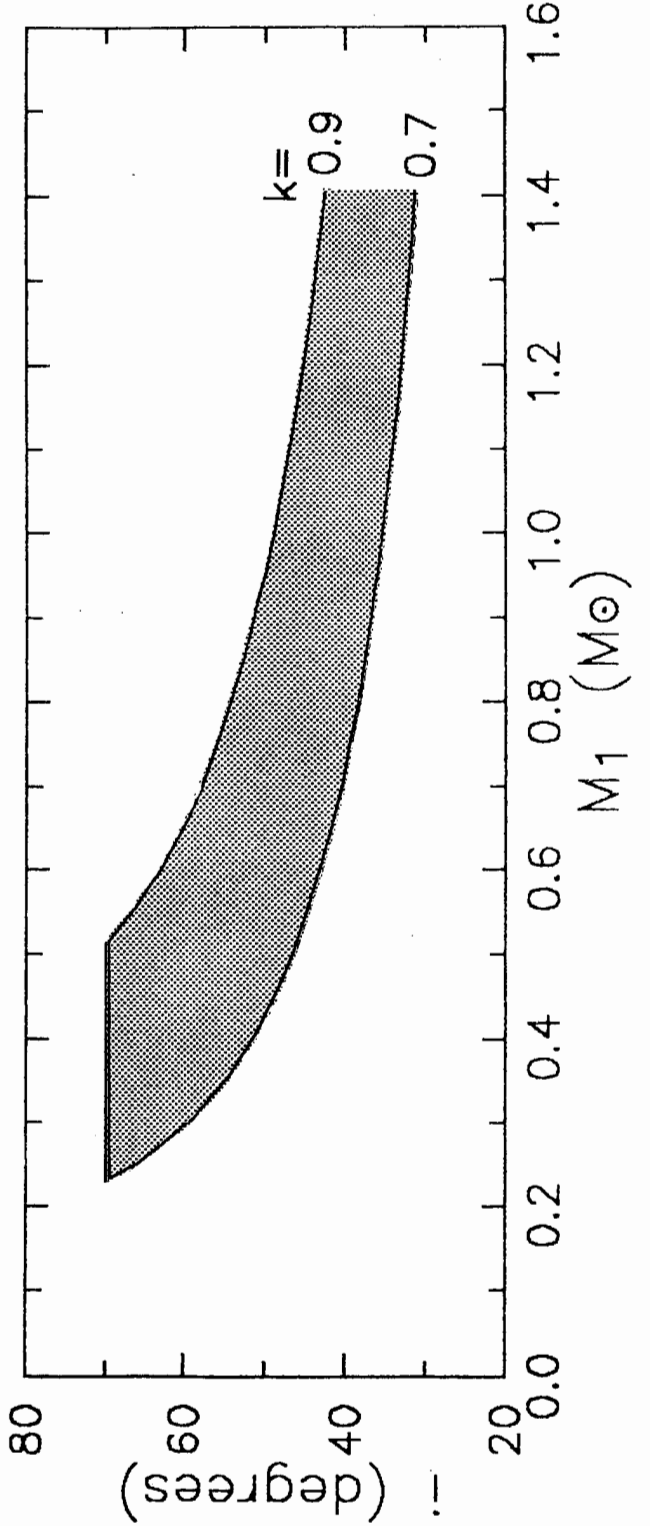

Fig. 7

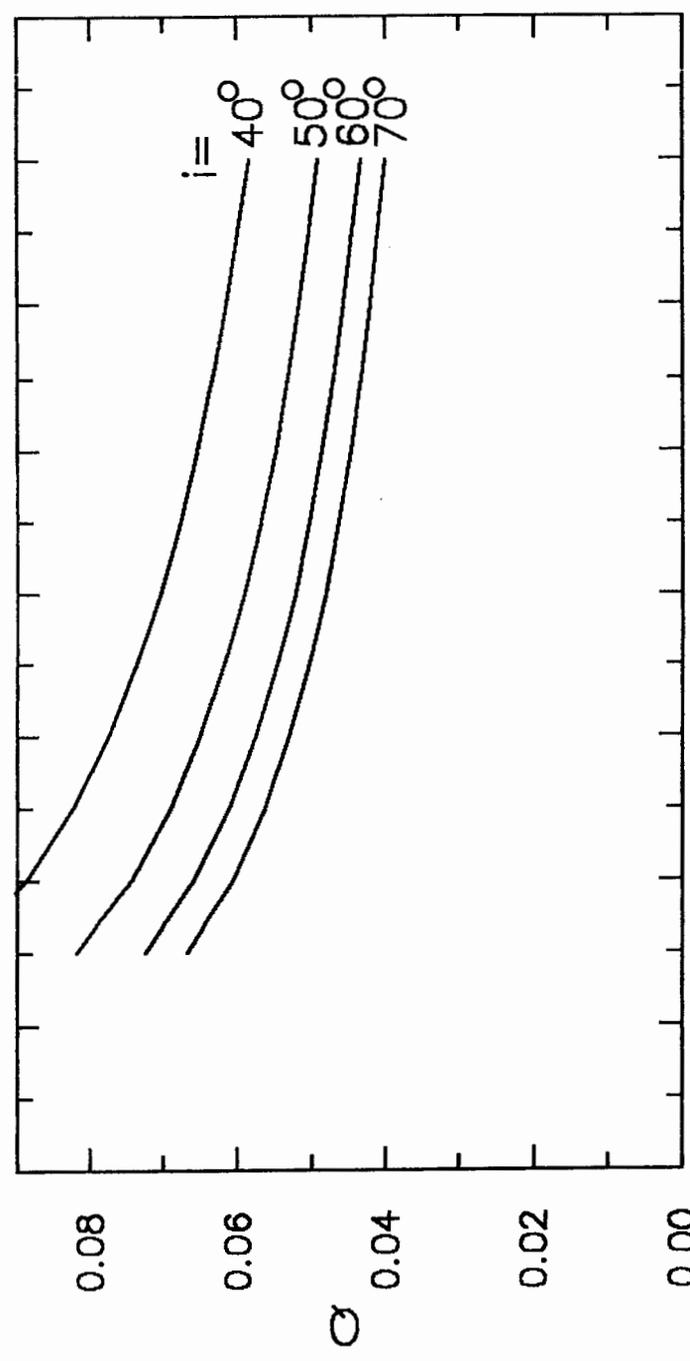
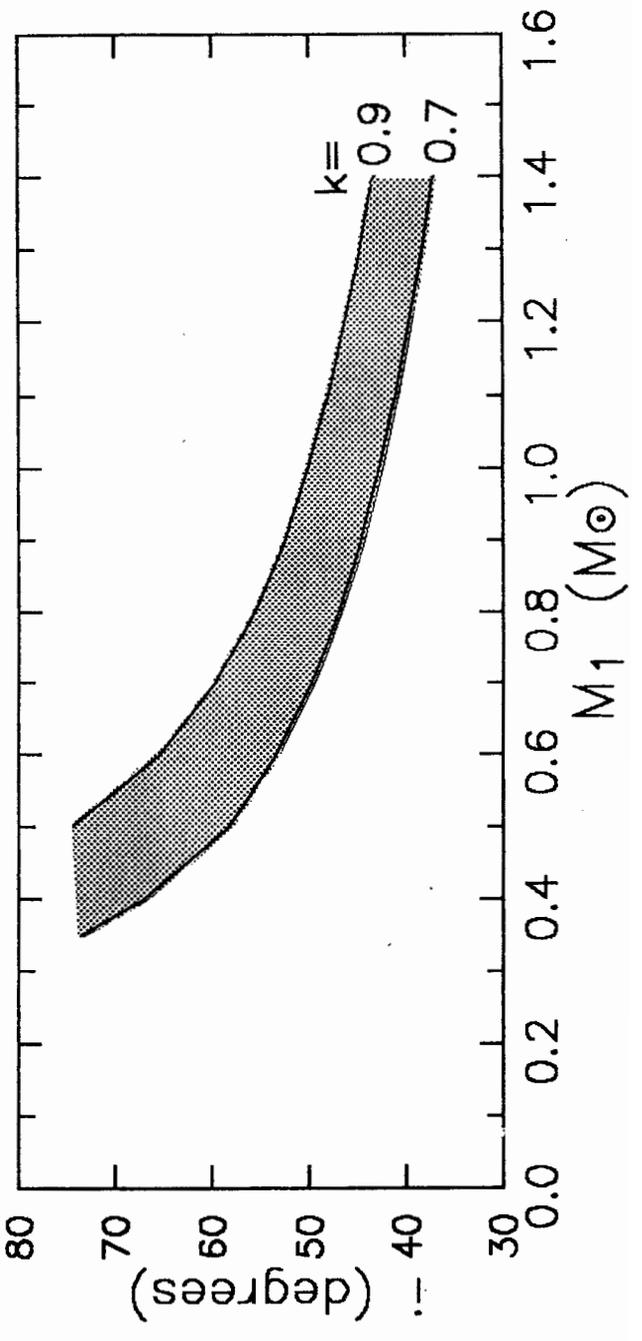

Fig 8

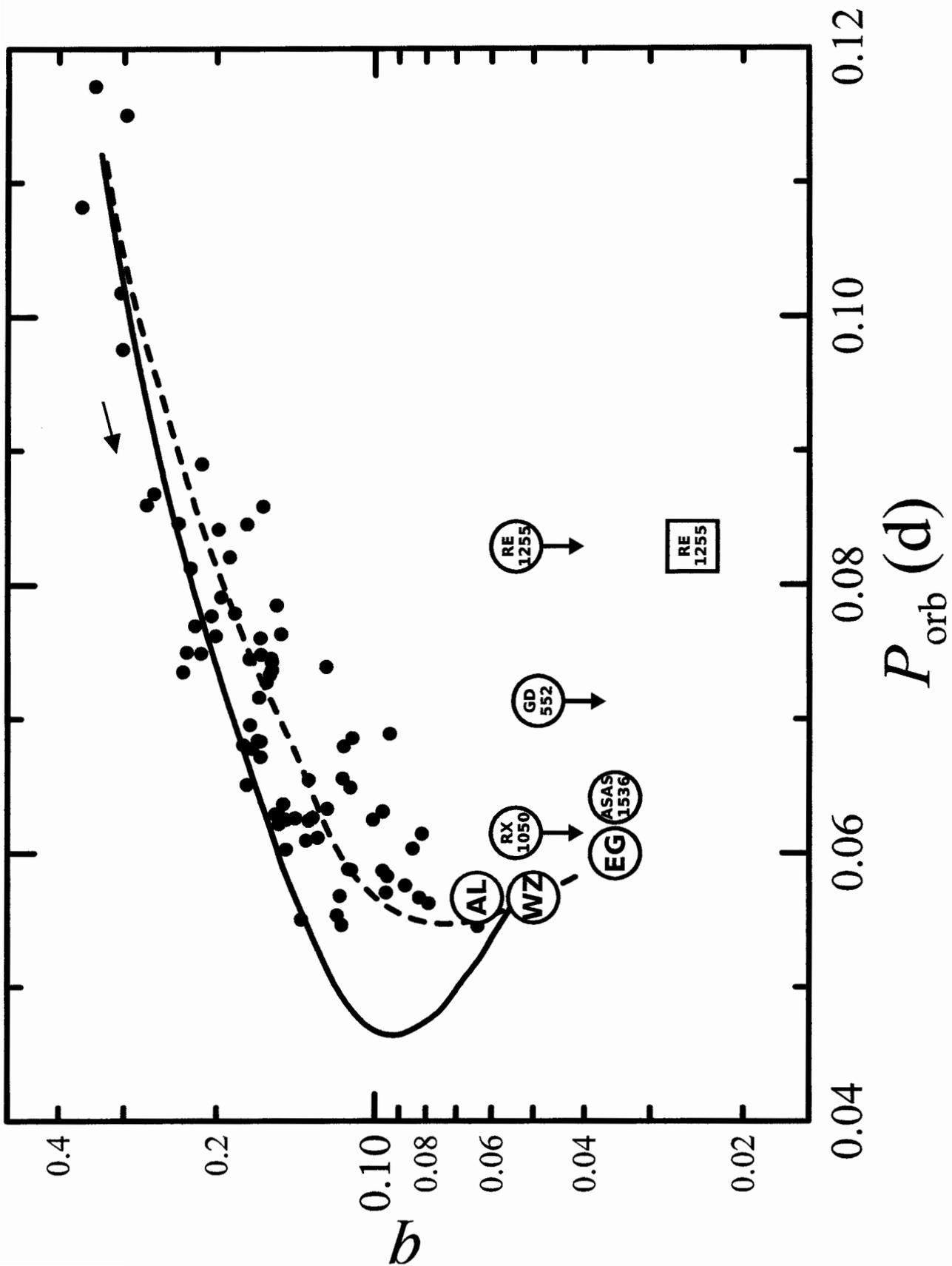

Fig 9

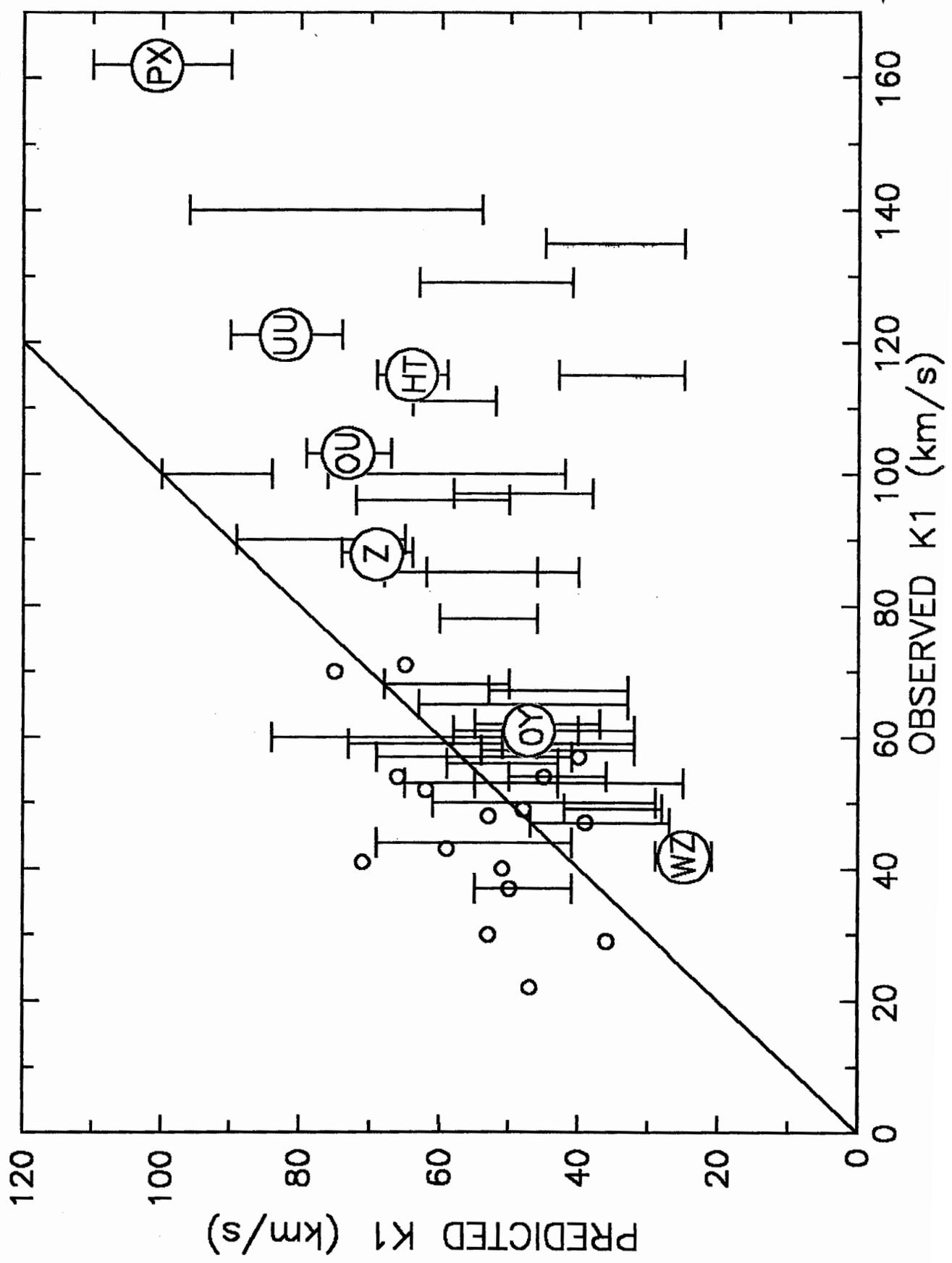

Fig A1